\documentclass[12pt]{article}
\usepackage{amsmath}

\usepackage[super,compress]{cite}
\usepackage{graphicx}
\usepackage{comment}
\renewcommand{\d}{{\rm d}}
\newcommand{\rl}{\ell}
\newcommand{\tilS}{\tilde{S}}
\newcommand{\bN}{\mbox{\boldmath$N$}}

\newcommand{\bpart}{\mbox{\boldmath$\partial$}}
\newcommand{\sgn}{\mathop{\rm sgn}\nolimits}
\renewcommand{\Re}{\mathop{\rm Re}\nolimits}
\renewcommand{\Im}{\mathop{\rm Im}\nolimits}
\newcommand{\bn}{\mbox{\boldmath$n$}}
\newcommand{\bom}{\mbox{\boldmath$m$}}
\newcommand{\bx}{\mbox{\boldmath$x$}}
\newcommand{\by}{\mbox{\boldmath$y$}}
\newcommand{\bl}{\mbox{\boldmath$l$}}
\newcommand{\rv}{\mbox{\rm v}}
\newcommand{\tl}{\tilde{l}}
\newcommand{\oDelta}{\overline{\Delta}}
\newcommand{\ophi}{\overline{\phi}}
\newcommand{\bq}{\mbox{\boldmath$q$}}
\newcommand{\bk}{\mbox{\boldmath$k$}}
\newcommand{\sh}{\mathop{\rm sh}\nolimits}
\newcommand{\ch}{\mathop{\rm ch}\nolimits}

\newcommand{\arsh}{\mathop{\rm arsh}\nolimits}

\allowdisplaybreaks[4]

\begin{document}

\title{On the discrete version of the Kerr–Newman solution
}

\author{V.M. Khatsymovsky \\
 {\em Budker Institute of Nuclear Physics} \\ {\em of Siberian Branch Russian Academy of Sciences} \\ {\em
 Novosibirsk,
 630090,
 Russia}
\\ {\em E-mail address: khatsym@gmail.com}}
\date{}
\maketitle
\begin{abstract}

This paper continues our work on black holes in the framework of the Regge calculus, where the discrete version (with a certain edge length scale $b$ proportional to the Planck scale) of the classical solution emerges as an optimal starting point for the perturbative expansion after functional integration over the connection, with the singularity resolved.

An interest in the present discrete Kerr-Newman type solution (with the parameter $a \gg b$) may be to check the classical prediction that the electromagnetic contribution to the metric and curvature on the singularity ring is (infinitely) greater than the contribution of the $\delta$-function-like mass distribution, no matter how small the electric charge is.

Here we encounter a kind of a discrete diagram technique, but with three-dimensional (static) diagrams and with only a few diagrams, although with modified (extended to complex coordinates) propagators.

The metric (curvature) in the vicinity of the former singularity ring is considered. The electromagnetic contribution does indeed have a relative factor that is infinite at $b \to 0$, but, taking into account some existing estimates of the upper bound on the electric charge of known substances, it is not so large for habitual bodies and can only be significant for practically non-rotating black holes.

\end{abstract}


MSC classes: 83C27; 83C57

keywords: Einstein theory of gravity; minisuperspace model; piecewise flat spacetime; Regge calculus; Kerr–Newman black hole

\section{Introduction}

Simplicial gravity, arising originally as Regge calculus \cite{Regge} (RC), replaces the analysis of a smooth space-time manifold of general relativity (GR) with an analysis of a piecewise flat manifold composed of flat 4-dimensional tetrahedra or 4-simplices. Such a manifold is characterized by a countable set of variables (edge lengths). Meanwhile, the aspect in which the continuum nature of GR poses a problem is the quantum aspect, in which GR is formally a non-renormalizable theory. Since piecewise flat manifolds can approximate a Riemannian manifold with arbitrarily high accuracy \cite{Fein,CMS}, RC can be used as a regularization when calculating gravitational path integrals \cite{HamWil1,HamWil2,Ham1}. The calculations can be simplified by narrowing the set of possible 4-simplices to a few types, which allows us to cover more complex cases while retaining essential degrees of freedom in the theory of Causal Dynamical Triangulations (CDT) \cite{cdt,cdt1}. An approach is also possible in which the space-time is assumed to be really piecewise flat \cite{Mik}.

Regge's name is also associated with the correspondence between the partition function of three-dimensional gravity over piecewise flat manifolds with some fixed (simplicial) boundary and the sum over such manifolds of products of 6j-symbols corresponding to tetrahedra from these manifolds with edge lengths that are just moments and are quantized as moments \cite{Pon}. Generalization of this construction to four dimensions leads to recent spin foam models of quantum gravity \cite{Per}.

As for black holes, RC was used as an approximation for the classical analysis of the Schwarzschild and Reissner–Nordström problems \cite{Wong}. Subsequently, for the analysis of classical problems, a more efficient discrete method was proposed \cite{Bre2}. At the quantum level, the Schwarzschild problem has been actively investigated in the framework of Loop Quantum Gravity (LQG), and the central singularity was resolved \cite{Ash1,Ash2}. Though LQG is not a discrete theory, the mechanism for eliminating the singularity lies in the discrete area spectrum and therefore in the presence of an area quantum, as in the lattice theory.

Taking into account some features of the discrete functional integral measure, we obtain a configuration with some finite nonzero typical edge length scale $b$ as an optimal starting point for the perturbative expansion of the functional integral. (There is an indirect analogy with a body spontaneously choosing an equilibrium position in a potential well.) In other words, the scale $b$ is dynamically defined inside simplicial gravity itself in quantum theory. It is proportional to the Planck length and thus to $\sqrt{ \hbar }$. In the classical limit $\hbar \to $, we have $b \to 0$, that is, passing to continuum. Thus, discreteness here is a quantum effect.

Also in RC, we have a unique case when the field variables, that is, the lengths, play at the same time the role of spacings of some lattice in space-time. Thus, one can also say that we have a {\it dynamical lattice} in which spacings are set (loosely fixed) internally as a result of the (quantum) dynamics of the system.

The physical motivation for applying RC to black holes is the possibility of resolving singularities in black holes inherent in the classical (continuum) solutions of the Einstein equations, due to the dynamical lattice effect, and the possibility of analyzing the metric and fields at (and in the vicinity of) the points where the curvature and/or fields are extremely large (usually these points are the former singularity points). This is achieved by passing to the simplicial realization of the space-time and to quantum theory, in which $b \neq 0$. The procedure and result of such a resolution of singularities are specific when we have a rotation (the former singularity points form an extended set in the form of a ring) and/or an electric charge (there is a bulk source of the gravity field due to the electromagnetic field in addition to the $\delta$-function-like source).

Mathematically, this analysis is motivated by the possibility of finding and studying the black hole type solutions of the Regge calculus with a (nonzero) typical edge length scale $b$ (although the division into physical and mathematical aspects of the issue is somewhat arbitrary). Or, in other words, the possibility of developing a simplicially discretized form of a black hole-type solution. A feature of such a coordinate-free description is that it does not have to suffer from coordinate-caused singularities such as horizons, only real physical singularities should show up. Our earlier results reduce the Regge skeleton equations to a finite-difference form of the Einstein equations, here simplified to a discrete Poisson equation. The specificity of this equation may consist in the presence of bulk sources due to the electromagnetic field and/or the ring shape of the set of singularity points of the solution of its continuum counterpart due to rotation. We have previously verified that a discrete rotating metric solution can be obtained by extending the corresponding discrete static solution to complex coordinates in a similar way that the continuum rotating (Kerr) metric solution can be obtained from the corresponding continuum static (Schwarzschild) solution. A similar situation will take place in the presence of a charge in the present paper.

For the discrete Schwarzschild problem, we consider such a configuration that is Schwarzschild-like at large distances (where discreteness can be neglected) \cite{Kha1,our}. The central singularity is cut off at the typical edge length scale. The simplicial electromagnetic field has been formulated in the literature \cite{Sor,Wein}. Taking this into account, we have analyzed the discrete Reissner–Nordström problem \cite{our3}. The analysis is reduced to the analysis of a finite-difference form of the Einstein-Maxwell equations. In addition to eliminating the central singularity, we obtain a refinement of the behavior of the electromagnetic contribution to the metric when approaching the center (an abrupt sign change).

Now consider the case when, in addition to the charge, there is also an essential rotation, i.e., a solution of the Kerr–Newman \cite{New,Ker} type here with the parameter $a >> b$. (We consider the uncharged case, Kerr geometry, in \cite{Kha2}.) An additional motivating reason for this analysis might be to test whether the electromagnetic part dominates the metric and curvature on the former singularity ring, no matter how small the charge is, as in the continuum theory. Here we discuss the metric (more exactly, its electromagnetic part), analyze where it experiences a sharp jump (in the vicinity of the former singularity ring), and get the first few terms in the expansion of the metric over coordinate variations from this ring. We briefly describe the approach in Section \ref{method}. For our purposes, it all comes down to solving a finite-difference form of the continuum Einstein-Maxwell equations with a certain lattice spacing $b$. The calculation is described in Section \ref{calculation}. Subsection \ref{integral} specifies the discrete equation to be solved for the metric and writes out an integral expression for its solution in quasi-momentum representation. The latter can be viewed as a combination of a few 1-loop and 3-loop discrete diagrams. The general diagram technique in simplicial gravity was considered in [\citen{HamLiu}]. In the diagrammatic aspect, if the simplicial subdivision of the hypercubic lattice \cite{RocWil} is used, which is the case here, the simplicial theory does not differ much from the finite-difference form of the continuum theory. Now our diagrams are 3-dimensional (static) ones with modified (extended to complex coordinates) propagators. Due to the latter modification in the integrand, we have oscillating exponents from expressions proportional to $a/b$ and having a saddle point, and we can perform integration over part of the variables using the saddle point method in Subsection \ref{saddle}. In Subsection \ref{expansion}, the expansion of the metric is found to be contributed by small (for small $b$) quasi-momenta for most terms (structures or monomials in coordinate shifts) and expressed by comparatively simple integrals; but just the first few (three) terms are contributed by the maximum quasi-momenta and require a more complicated integration.

\section{The method}\label{method}

The RC strategy involves summing or averaging over all simplicial structures, which from a numerical point of view is the most difficult problem qualitatively, and one can hope to find some simplifying analytic theorems and methods; here, as already mentioned, some fixed (periodic) structure will be used.

The edge lengths are a priori not forbidden to be arbitrarily small with a significant probability, which reproduces the continuum theory. For the presence of a cut off effect in RC, a mechanism is needed for the emergence of a finite nonzero typical edge length scale, which we can obtain due to some features of the discrete functional integral measure \cite{our1}. If we want to use some fundamental principles for constructing the functional integral, such as the canonical Hamiltonian formalism, we must be able to analyze the system in a continuous time limit. In RC, the continuous time limit is not well-defined if the system is described solely by edge length variables. The way out is to extend the set of variables. Using the connection variables introduced by Fr\"{o}hlich \cite{Fro}, we can write the RC action in terms of tetrad variables (edge vectors in local Minkowski frames in 4-simplices) together with independent $SO(3,1)$ rotations, more exactly, as a sum of contributions from their (independent) self-dual and anti-self-dual parts \cite{Kha}. (Excluding the connection variables at the classical level leads to the original RC action.)

In the tetrad-connection variables, we can consistently pass to the continuous time limit, when the lengths (of the projections) of the edges in a certain direction, taken as a time direction, are made arbitrarily small, construct the Hamiltonian formalism, and canonically quantize the system. In this formalism, certain bivectors and connection matrices are canonically conjugate. We can write the result of this quantization as a functional integral. But this is for those piecewise flat geometries that are infinitely close to smooth ones in the above time direction. If we need a functional integral on the general piecewise flat space-time, then it is natural to require that it goes into the found continuous time form, whichever direction is chosen as the direction of time with the corresponding smoothing of the geometry along it by taking the lengths (of the projections) of the edges arbitrarily small along it. This requirement is rather restrictive, and it is not a priori obvious that such a functional integral exists.

However, it exists in an extended superspace of independent area tensors, that is, those that do not have to correspond to any real set of edge vectors. A functional integral can be viewed as a linear functional on the superspace (of functions) of observables. The usual superspace of observables that are functions of bivectors constructed from vectors of real edges can be embedded in the extended superspace of independent area tensors by multiplying by a $\delta$-functional factor that takes into account the existence of a defining set of edge vectors for the current set of area tensors. This mapping of superspaces induces its dual mapping of linear functionals on them. Thus, it gives the desired functional integral on the actual observables of interest to us.

The $\delta$-function factor ensures that the conditions for the existence of a defining set of edge vectors in {\it separate} 4-simplices (as well as the conditions for the continuity of the metric induced on 3-faces through these faces) are satisfied. It can behave as a power of the volume of a 4-simplex under arbitrary deformations of this 4-simplex. In the continuum limit, this would mean behavior as a power of $\det \| g_{\lambda \mu} \|$, i.e., as a scalar density. There is no strict requirement that the measure be, say, a scalar rather than a scalar density (as for the continuity conditions, the corresponding part of the $\delta$-function factor can be fixed by the requirement that it be invariant under arbitrary deformations of the 3-faces, leaving them in the same 3-planes). Therefore, we take as free some parameter $\eta$ whose change by $\Delta \eta$ corresponds to the additional factors in the measure $V^{\Delta \eta}_{\sigma^4}$ with the 4-volume $V_{\sigma^4}$ of the 4-simplices $\sigma^4$ or $ ( - \det \| g_{\lambda \mu} \| )^{\Delta \eta / 2}$ in the continuum theory.

In the functional integral with the above RC action $S_{\rm g}(\rl , \Omega )$ in terms of edge variables $\rl$ and connection variables $\Omega$ and the measure $\d \mu ( \rl ) {\cal D} \Omega$, we can integrate over ${\cal D} \Omega$ (the invariant measure) and obtain some resultant phase $\tilS_{\rm g} ( \rl )$ and measure $F ( \rl ) D \rl$,
\begin{equation}\label{int-S D-Omega=F}                                     
\int \exp [ i S_{\rm g}(\rl , \Omega  ) ] ( \cdot ) \d \mu ( \rl ) {\cal D} \Omega = \int \exp [ i \tilS_{\rm g} ( \rl ) ] ( \cdot ) F ( \rl ) D \rl .
\end{equation}

\noindent These phase and measure are the argument and modulus of a complex value, the result of this integration of interest to us, the calculation of which in closed form requires the use of some expansion of $S_{\rm g}(\rl , \Omega )$. For accuracy, it is important that the characteristic under study (argument or modulus) would receive the main contribution from the leading term of such an expansion used.

Thus, for the phase $\tilS_{\rm g} ( \rl )$ it is appropriate to expand $S_{\rm g}(\rl , \Omega )$ over variations of $\Omega$ from the background $\Omega = \Omega_0 ( \rl )$ (for example, over $\omega \in$ so(3,1) such that $\Omega = \Omega_0 \exp \omega$). Since the equations of motion for $\Omega$ are fulfilled at $\Omega = \Omega_0$, the term $\propto \omega$ is equal to zero. The bilinear term leads to a Gaussian integral in which typical values of $\omega$ are $l_{\rm Pl} / l$, where $l$ is a typical value of the edge lengths. Looking ahead, we will replace $l$ with $b$. The expansion goes in powers of $l_{\rm Pl} / b$, which is a small parameter for $b \gg l_{\rm Pl}$. The zeroth order term $S_{\rm g}(\rl , \Omega_0 ( \rl ) )$ in the phase $\tilS_{\rm g} ( \rl )$ is the RC action $S_{\rm g} ( \rl )$ by the definition of the connection representation $S_{\rm g}(\rl , \Omega )$. A similar integration over the connection in the functional integral of the continuum theory in terms of the tetrad-connection variables gives exactly the functional integral in terms of purely metric variables and the Einstein action for the phase. Now, in the discrete theory, we start with the above "Gaussian approximation" and have the phase $\tilS_{\rm g} ( \rl )$ expanded over $( l_{\rm Pl} / b )^2$ with the RC action $S_{\rm g} ( \rl )$ as the leading order term.

As for the measure $F ( \rl )$, the expansion over discrete analogs of the ADM lapse-shift functions \cite{ADM1} $(N, \bN )$ (certain edge vectors) allows to capture the main part of the effect already in the zeroth order. Here we start from a "factorization approximation", since in the zeroth order only spatial and diagonal triangles contribute to the action $S_{\rm g}(\rl , \Omega )$, but not temporal triangles. These triangles form some (maximal) set on which the matrices of holonomy $R$ of $\Omega$ or the curvature matrices (on which $S_{\rm g}(\rl , \Omega )$ depends) can be taken as independent connection variables.
Then the connection integration can be performed for each triangle separately.

In the presence of an electromagnetic field described by a set of simplicial variables $A$, we proceed from the full action
\begin{equation}                                                            
S(\rl , \Omega , A ) = S_{\rm g}(\rl , \Omega  ) + S_{\rm em}(\rl , A  )
\end{equation}

\noindent specified by us earlier \cite{our3}. The process of integration over $\Omega$ is not affected and replaces the measure $\d \mu ( \rl ) {\cal D} \Omega$ with $F ( \rl ) D \rl$, the phase $S_{\rm g}(\rl , \Omega )$ with $\tilS_{\rm g} ( \rl ) \approx S_{\rm g} ( \rl )$ and $S(\rl , \Omega , A )$ with $S(\rl , A ) = \tilS_{\rm g} ( \rl ) + S_{\rm em}(\rl , A  ) \approx S_{\rm g} ( \rl ) + S_{\rm em}(\rl , A  )$.

Suppose we have passed from $\rl = (l_1, \dots, l_n )$ to some new variables $u = (u_1, \dots, u_n )$ that make the measure $F ( \rl ) D \rl$ Lebesgue $D u$ and have expanded $S(\rl , A )$ in a neighborhood of some starting point $\rl_{(0)} = \rl ( u_{(0)} )$, considering $A$ first as parameters,
\begin{equation}                                                            
S (\rl , A ) = S (\rl_{(0)} , A ) + \frac{1}{2} \sum_{j, k, l, m} \frac{\partial^2 S (\rl_{(0)} , A )}{\partial l_j \partial l_l} \frac{\partial l_j (u_{(0)} )}{\partial u_k} \frac{\partial l_l (u_{(0)} )}{\partial u_m} \Delta u_k \Delta  u_m + \dots ,
\end{equation}

\noindent where $\Delta u = u - u_{(0)}$ and
\begin{equation}\label{dS/dl=0}                                             
\frac{\partial S(\rl_{(0)} , A )}{\partial l_j} = 0 .
\end{equation}

\noindent Just as we impose the extremum condition (\ref{dS/dl=0}) on the zero-order term to maximize the contribution, now we can also impose a minimum condition on the determinant of the second-order form in the exponential for this, that is, the following maximization condition,
\begin{equation}\label{def-l0}                                              
F (\rl_{(0)} )^2 \det \left \| \frac{\partial^2 S (\rl_{(0)} , A )}{\partial l_i \partial l_k} \right \|^{-1} = \mbox{ maximum}.
\end{equation}

In the measure $F ( \rl ) D \rl$, one can distinguish the dependence on the length (or area) scale in the domain where such a scale can be introduced. It is more natural to write this in terms of a typical area scale $\rv$, from which the length scale $l$ follows from $\rv = l^2 / 2$. This dependence looks like $[f ( \rv ) ]^T \rv^{- 1} \d \rv$, where $f ( \rv ) = \rv^{2 ( \eta - 5 ) / 3 } \exp{ ( - \pi \rv l_{\rm Pl}^{- 2} ) }$ with the aforementioned $\eta $. Here $T$ is the number of certain - spatial and diagonal - triangles in the domain, analogues of infinitesimal areas built on pairs of spatial (according to the world index) tetrad vectors in the continuum theory. The actions $S_{\rm g} ( \rl )$ and $S_{\rm em}(\rl , A  )$ depend on $\rv$ as $\rv^1$ and $\rv^0$, respectively. Thus, depending on whether $S_{\rm g} ( \rl )$ or $S_{\rm em}(\rl , A  )$ dominates the sum $S(\rl , A )$, equation (\ref{def-l0}) (where the set $\rl$ constitutes one variable $\rv$) requires $[f ( \rv ) ]^T \rv^{- 1 / 2 }$ or $[f ( \rv ) ]^T $ to be maximized, respectively. The maxima of these expressions are located at $2 \pi l_{\rm Pl}^{- 2} \rv = 4 ( \eta - 5 ) / 3 - 1 / T$ and $2 \pi l_{\rm Pl}^{- 2} \rv = 4 ( \eta - 5 ) / 3$, respectively. With the assumed greatness of the length scale $b \gg l_{\rm Pl}$ and any significant $T \gg 1$, suitable for defining the concept of a typical length scale, the impact of the presence of $S_{\rm em}(\rl , A  )$ is not significant. From here, the length scale itself is obtained,
\begin{equation}\label{b=sqrt}                                              
b = l_{\rm Pl} \sqrt{ 4 ( \eta - 5) / (3 \pi ) } , ~~~ l_{\rm Pl} = \sqrt{8 \pi G} .
\end{equation}

As for $A$, it is also necessary to expand $S (\rl , A )$ over it and write down the equation of motion for it,
\begin{equation}\label{dS/dA=0}                                             
\frac{\partial S(\rl_{(0)} , A_{(0)} )}{\partial A} = 0 ,
\end{equation}

\noindent at the starting point $\rl_{(0)} , A_{(0)}$. In overall, the equations of motion (\ref{dS/dl=0}) (at $A = A_{(0)}$) and (\ref{dS/dA=0}) and the maximization condition (\ref{def-l0}) (at $A = A_{(0)}$) define the starting point.

In calculations, we consider approximate estimates when approaching extremal (which are close to the former singular) points from the near-continuum case, when the metric/field variations from simplex to simplex in the system configuration $\rl_{(0)} , A_{(0)}$ are small, so that we can work in leading order over these variations. The action in the neighborhood of this point will essentially depend on a smaller set of variables or their combinations than the complete simplicial set $\rl , A$ (in fact, this will be a finite-difference form of the continuum action). The above considerations about the extremum of the exponent lead to the equations of motion (\ref{dS/dl=0}) (at $A = A_{(0)}$), (\ref{dS/dA=0}), where now we should mean by $\rl , A$ this smaller set of variables.

We are faced with the necessity to solve the simplicial Einstein-Maxwell equations.

For the gravitational part, we have considered \cite{our2} a simplicial complex and defined a piecewise constant metric on it by assigning (generally speaking, freely) certain coordinates to the vertices. Discrete Christoffel symbols were defined, the defect angles were expressed in terms of them and the RC action found. This is adapted for the expansion over metric variations from 4-simplex to 4-simplex. The connection with the continuum notations is best manifested on a periodic simplicial complex. Using the simplest such complex, consisting of 4-cubes, each of which is divided by diagonals into 4!=24 4-simplices  \cite{RocWil}, we have a finite-difference form of the Hilbert-Einstein action for the main term,
\begin{eqnarray}\label{DM+MM}                                               
S_{\rm g} ( \rl ) &  = & \frac{1}{16 \pi G}\sum_{\rm 4-cubes} {\cal K}^{\lambda \mu}{}_{\lambda \mu} \sqrt{g} , ~~ {\cal K}^\lambda{}_{\mu \nu \rho} \! = \! \Delta_\nu M^\lambda_{\rho \mu} \! - \! \Delta_\rho M^\lambda_{\nu \mu} \! + \! M^\lambda_{\nu \sigma} M^\sigma_{\rho \mu} \! - \! M^\lambda_{\rho \sigma} M^\sigma_{\nu \mu} , \nonumber \\ \hspace{0mm}
M^\lambda_{\mu \nu} & = & \frac{1}{2} g^{\lambda \rho} (\Delta_\nu g_{\mu \rho} + \Delta_\mu g_{\rho \nu} - \Delta_\rho g_{\mu \nu}), ~~~ \Delta_\lambda = 1 - \overline{T}_\lambda . \hspace{0mm}
\end{eqnarray}

\noindent where the shift operator acts as $T_\lambda f(\dots , x^\lambda , \dots ) = f(\dots , x^\lambda + 1 , \dots )$ for a function $f$; $\overline{T}_\lambda$ is its Hermitean conjugate, coinciding with $T^{-1}_\lambda$.

As for the simplicial electromagnetic action, using the Sorkin-Weingarten formulation \cite{Sor,Wein} within our approach, we again have a finite-difference form of the continuum electromagnetic action \cite{our3}.

In practical use of the expansion of the measure $F ( \rl ) D \rl$ over the discrete lapse-shift functions, the obstacles may be the points of strong growth of these functions. The way out may be to use a reference frame close to the synchronous frame, the construction of which is always possible \cite{Wald}. Then the interpolating metric in the finite-difference form (\ref{DM+MM}) has $(N, \bN ) = (1, {\bf 0})$. We work in the leading order over metric variations, and in this order, finite differences follow the same rules as ordinary derivatives. (In principle, it is possible to calculate and add to (\ref{DM+MM}) non-leading orders over metric variations, but they are cumbersome, lattice-specific, and are not described exclusively in terms of $g_{\lambda \mu}$ at each vertex.) In particular, in the finite-difference form of the Hilbert-Einstein action, we can go from a metric close to the Kerr–Newman metric in a synchronous frame of reference to a metric close to the Kerr–Newman metric in some other frame of reference (a kind of diffeomorphism invariance on the discrete level). As the latter frame of reference, we can consider the Kerr-Schild coordinate system.

Thus, the aim is to analyze the solution to a finite-difference form of the continuum Einstein-Maxwell equations with a certain lattice spacing $b$ close to the continuum Kerr–Newman solution in the Kerr-Schild coordinates \cite{Ker}.

\section{Calculation}\label{calculation}

\subsection{Integral Expression For Metric Function}\label{integral}

The continuum solution takes the form ($\lambda$, $\mu$, $\nu$, \dots = 0, 1, 2, 3)
\begin{equation}\label{A,ds}                                                
A_\lambda = - A_0 n_\lambda , ~~~ l_\lambda = - l_0 n_\lambda , ~~~ \d s^2 = - \d \tau^2 + \d x_1^2 + \d x_2^2 + \d x_3^2 + (l_\lambda \d x^\lambda )^2 ,
\end{equation}

\noindent where
\begin{eqnarray}\label{n,l,A0}                                             
n_\lambda = \left ( -1, \frac{r x_1 - a x_2}{r^2 + a^2}, \frac{r x_2 + a x_1}{r^2 + a^2}, \frac{x_3 }{r } \right ), ~~~ l_0^2 = ( l_0^2 )_g + ( l_0^2 )_{em}, \nonumber \\ \frac{A_0 }{Q } = \frac{( l_0^2 )_g }{r_g } = \frac{r }{r^2 + a^2 \cos^2 \theta }, ~~~ ( l_0^2 )_{em} = - \frac{Q^2 }{r^2 + a^2 \cos^2 \theta }
\end{eqnarray}

\noindent and $r, \theta$ are related to $x_1, x_2, x_3$ according to
\begin{equation}                                                           
r \cos \theta = x_3, ~~~ r^4 - r^2 ( x_1^2 + x_2^2 + x_3^2 - a^2 ) - a^2 x_3^2 = 0 .
\end{equation}

In the leading order $[l ]^2$, we have for the Ricci tensor ($k$, $l$, $m$, \dots = 1, 2, 3)
\begin{eqnarray}\label{R=ddll}                                             
& & 2 R_{00} = - \bpart^2 ( l_0^2 ) , \nonumber \\
& & 2 R_{0k} = - \bpart^2 ( l_0 l_k ) + \partial_k \partial_m ( l_0 l_m ) , \nonumber \\
& & 2 R_{kl} = - \bpart^2 ( l_k l_l ) + \partial_k \partial_m ( l_l l_m ) + \partial_l \partial_m ( l_k l_m ) .
\end{eqnarray}

The electromagnetic potential as given by (\ref{A,ds}, \ref{n,l,A0}) can be written as
\begin{equation}                                                           
A_\lambda = \tl_0 \tl_\lambda \sgn A_0 ,
\end{equation}

\noindent where $\tl_\lambda$ is collinear to $l_\lambda$. It defines some Kerr metric and thus it is a solution to the vacuum Einstein equations for the algebraically special $g_{\lambda \mu}$. As such, it obeys the relation \cite{Chandra}
\begin{equation}                                                           
\tl^\nu \partial _\nu \tl_\lambda = \psi (\bx ) \tl_\lambda ,
\end{equation}

\noindent where $\psi ( \bx )$ is a function. Using it simplifies the procedure of raising/lowering the indices of the electromagnetic tensor,
\begin{equation}                                                           
F^{\lambda \mu} = \eta^{\lambda \nu} \eta^{\mu \rho} F_{\nu \rho}, ~~~ F_{\lambda \mu} = \partial_\lambda A_\mu  - \partial_\mu A_\lambda .
\end{equation}

\noindent Then
\begin{eqnarray}\label{dF=ddll}                                            
& & F^{\lambda 0 }{}_{; \lambda } = - \bpart^2 ( \tl_0^2 ) , \nonumber \\
& & F^{\lambda k }{}_{; \lambda } = \bpart^2 ( \tl_0 \tl_k ) - \bpart_k \bpart_m ( \tl_0 \tl_m ) .
\end{eqnarray}

It is seen that $R_{00}$, $R_{0k}$ (\ref{R=ddll}) when $l_\lambda$ is replaced by $\tl_\lambda$ turn out to be proportional to $F^{\lambda 0 }{}_{; \lambda }$, $F^{\lambda k }{}_{; \lambda }$ (\ref{dF=ddll}), respectively. Thus, the validity of the vacuum Einstein 00- and 0k- equations for $\tl_\lambda$ means the validity of the Maxwell equations for $A_\lambda$. In particular, the solution of the Kerr type to the 0-component $F^{\lambda 0 }{}_{; \lambda } = 0$ of the Maxwell equations satisfies their k-components, where the 4-vector $n_\lambda$ is that one given for the Kerr solution.

The Kerr metric $g_{(g ) \lambda \mu} = ( l_0^2 )_g n_\lambda n_\mu$ satisfies the vacuum Einstein equations. The actual metric satisfies the inhomogeneous equations $R_{\lambda \mu} = 8 \pi G ( T_{\lambda \mu} - g_{\lambda \mu} T / 2 ) = 8 \pi G T_{\lambda \mu} $ with
\begin{eqnarray}\label{4piTlm}                                             
& & 4 \pi T_{0 0} = ( \partial_k A_0 )^2 - \frac{1}{2} (n_k \partial_k A_0 )^2 + 2 \frac{a^2 }{\rho^4 } n_3^2 A_0^2 , \nonumber \\
& & 4 \pi T_{0 k} = n_k ( \partial_m A_0 )^2 - ( \partial_k A_0 ) ( n_m \partial_m A_0 ) - 2 \frac{a }{\rho^2 } n_3 \epsilon_{k l m} (\partial_l A_0 ) n_m A_0 , \nonumber \\
& & 4 \pi T_{k l} = n_k n_l ( \partial_m A_0 )^2 - ( n_m \partial_m A_0 ) ( n_k \partial_l A_0 + n_l \partial_k A_0 ) \nonumber \\ & & \phantom{4 \pi T_{k l} =} - 2 \frac{a }{\rho^2 } n_3 A_0 (n_k \epsilon_{l m q} + n_k \epsilon_{k m q} ) ( \partial_m A_0 ) n_q + 4 \frac{a^2 }{\rho^4 } n_3^2 A_0^2 ( \delta_{k l} - n_k n_l ) \nonumber \\ & & \phantom{4 \pi T_{k l} =} + \frac{1}{2} ( n_m \partial_m A_0 )^2 \delta_{k l} - 2 \frac{a^2 }{\rho^4 } n_3^2 A_0^2 \delta_{k l} .
\end{eqnarray}

In particular, the solution $l_0^2$ (or, in fact, $( l_0^2 )_{em}$) of the 00-component $R_{0 0} = 8 \pi G T_{0 0}$ satisfies the 0k- and kl-components, where the 4-vector $n_\lambda $ is given.

In the continuum theory, there is a simple identity, according to which the scalar potential of the electromagnetic field and $(l_0^2)_g$ are the real part of the Coulomb/Newton potential, extended to complex coordinates,
\begin{equation}\label{l02g=A0=Rephi}                                      
\frac{(l_0^2)_g}{r_g} = \frac{A_0}{Q} = \Re \phi (\bx + i \by) , ~~~ \phi (\bx ) = \frac{1 }{\sqrt{\bx^2}} , ~~~ \by = (0, 0, y_3) , ~~~ y_3 = a \sgn y_3 .
\end{equation}

\noindent This allows us to suggest that the discrete solution for $(l_0^2)_g$ and $A_0$ is obtained by taking the real part of the extension of the {\it discrete} non-rotating solution to complex coordinates. Indeed, we found \cite{Kha2} that this assumption turned out to be true for the Kerr solution, where $r_g^{-1} l_0^2 = \Re \phi (\bx + i \by)$ with the discrete Coulomb/Newton potential $\phi (\bx )$ indeed satisfies Einstein's equation (discrete Poisson's equation) with a $\delta$-function-like source on the RHS supported on a disk close to the disk formed by the former singularity ring. In the considered charged case, we have the same discrete equations for $r_g^{-1} (l_0^2)_g$ and $Q^{-1} A_0$, and $(l_0^2)_g$ and $A_0$ are obtained by the same extension to complex coordinates (\ref{l02g=A0=Rephi}) from the discrete Coulomb/Newton potential $\phi (\bx )$.

Note that (\ref{l02g=A0=Rephi}) make up a (smaller, simpler, and more obvious) part of the relationship between continuous rotating and non-rotating solutions, the use of which is known as the Newman-Janis trick \cite{Newman,Rajan}.

In the discrete formalism, we need to rewrite the explicit dependence on the (continuum) coordinates in (\ref{4piTlm}) in the discrete form. Especially ambiguous is rewriting $n_3 \rho^{-2}$, $n_3^2 \rho^{-4}$, because these functions strongly vary in the vicinity of the singularity ring. This coordinate dependence is a remnant of using a specific metric field ansatz, whereas originally we have a common metric and field. Now we can identically (on the considered Kerr-Newman solution) transform the expressions of interest to the form containing only $\phi$ and no foreign coordinate dependence. In particular, we can substitute
\begin{equation}                                                           
\frac{a n_3 }{ \rho^2 } \sgn y_3 = - \Im \phi , ~~~ \frac{a^2 n_3^2 }{ \rho^4 } = ( \Im \phi )^2 .
\end{equation}

\noindent In $T_{00}$, this brings the third term to the form proportional to $( \Re \phi )^2 ( \Im \phi )^2$. In general, the result of calculating the expressions of interest to us for a specified $\phi$, due to axial symmetry, essentially depends on two variables, say, $r$ and $\cos \theta$, and can be uniquely written as a function of two variables, $ \Re \phi $ and $ \Im \phi $ or $\phi$ and $\overline{ \phi }$. As a result, the second plus the third terms in $T_{00}$ on the continuum solution take the form
\begin{eqnarray}                                                           
- \frac{1}{2} ( \bn \bpart \Re \phi )^2 + 2 \frac{a^2 }{\rho^4 } n_3^2 ( \Re \phi )^2 & = & - \frac{1}{2} ( \Re \phi )^4 + 3 ( \Re \phi )^2 ( \Im \phi )^2 - \frac{1}{2} ( \Im \phi )^4 \nonumber \\ & = & - \frac{1}{2} \Re ( \phi^4 ) .
\end{eqnarray}

In the discrete formalism, derivatives are substituted by finite differences. To preserve the symmetries as much as possible, we use a symmetric finite difference in $T_{\lambda \mu}$,
\begin{equation}                                                           
\partial_\lambda \Rightarrow \frac{1}{2} b^{-1} ( \Delta_\lambda - \oDelta_\lambda ) .
\end{equation}

In order $[l]^2$, the 00-component of the Einstein equations gives directly $( l_0^2 )_{em}$,
\begin{eqnarray}\label{DDl0^2em=(Dphi)^2}                                  
& & \sum^3_{j = 1} \oDelta_j \Delta_j ( l_0^2 )_{em} = 4 G Q^2 \left \{ \frac{1}{4} \left [ \sum^3_{j = 1} (\Delta_j - \oDelta_j ) \Re \phi \right ]^2 - \frac{1}{2} b^2 \Re (\phi^4 ) \right \} \nonumber \\ & & = 2 G Q^2 \Re \left \{ \frac{1}{4} \left [ \sum_j (\Delta_j - \oDelta_j ) \phi \right ]^2 + \frac{1}{4} \left [ \sum_j (\Delta_j - \oDelta_j ) \phi \right ] \left [ \sum_j (\Delta_j - \oDelta_j ) \overline{ \phi } \right ] \right. \nonumber \\ & & \left. \phantom{\sum_j} - b^2 \phi^4 \right \} .
\end{eqnarray}

\noindent Here $\phi$ follows by extending the discrete Coulomb/Newton potential $\phi_0 ( \bx )$ to complex coordinates,
\begin{equation}                                                           
\phi ( \bx ) = \phi_0 ( \bx + i \by ), ~~~ \by = (0, 0, a \sgn y_3 ) .
\end{equation}

\noindent The function $\phi_0 ( \bx )$ is defined as the solution of the finite-difference Poisson equation with a source at $\bx = {\bf 0}$ which has the continuum potential as a large distance asymptotic,
\begin{equation}                                                           
\sum^3_{j = 1} \oDelta_j \Delta_j \phi_0 ( \bx ) = 0 \mbox{ at } \bx \neq {\bf 0}, ~~~ \phi_0 ( \bx ) \to \frac{1}{\sqrt{\bx^2}} \mbox{ at } \bx \to \infty .
\end{equation}

\noindent We have found \cite{Kha2}
\begin{equation}                                                           
\phi ( \bx ) = \int_{- \pi }^\pi \int_{- \pi }^\pi \frac{ \d q_1 \d q_2 }{2 \pi b \sh \kappa } \exp \left \{ i \frac{ s }{ b } \left [ q_1 x_1 + q_2 x_2 - a \kappa \right ] - \frac{ | x_3 | }{ b } \kappa \right \}
\end{equation}

\noindent where
\begin{equation}                                                           
\ch \kappa \equiv \ch \kappa (q_1 , q_2 ) = 3 - \cos q_1 - \cos q_2 , ~~~ s \equiv \sgn (x_3 y_3 ).
\end{equation}

\noindent Then $2 \Re \phi = \phi |_{s = +1} + \phi |_{s = -1}$, $2 i \Im \phi = ( \phi |_{s = +1} - \phi |_{s = -1} ) s$. But, upon redefining $q_j \Rightarrow q_j s$,
\begin{equation}                                                           
\phi ( \bx ) = \int_{- \pi }^\pi \int_{- \pi }^\pi \frac{ \d q_1 \d q_2 }{2 \pi b \sh \kappa } \exp \left ( i \frac{ q_1 x_1 + q_2 x_2 }{ b } - \frac{ | x_3 | }{ b } \kappa - i s \frac{a}{b} \kappa\right ) ,
\end{equation}

\noindent therefore
\begin{equation}                                                           
\left. \begin{array}{r} \Re \phi \\ \Im \phi \end{array} \right \} = \int_{- \pi }^\pi \int_{- \pi }^\pi \frac{ \d q_1 \d q_2 }{2 \pi b \sh \kappa } \exp \left ( i \frac{ q_1 x_1 + q_2 x_2 }{ b } - \frac{ | x_3 | }{ b } \kappa \right ) \cdot \left \{ \begin{array}{l} \cos \left ( \frac{a}{b} \kappa \right ) \\ - s \sin \left ( \frac{a}{b} \kappa \right ) \end{array} \right. .
\end{equation}

\noindent Since $\Im \phi$ occurs on the RHS of the equation for $( l_0^2 )_{em}$ (\ref{DDl0^2em=(Dphi)^2}) being squared, it makes no difference which $s$ is taken, and we choose $s = 1$ for definiteness. Then
\begin{equation}\label{phi,ophi}                                           
\left. \begin{array}{r} \phi  \\ \overline{\phi } \end{array} \right \} = \int_{- \pi }^\pi \int_{- \pi }^\pi \frac{ \d q_1 \d q_2 }{2 \pi b \sh \kappa } \exp \left ( i \frac{ q_1 x_1 + q_2 x_2 }{ b } - \frac{ | x_3 | }{ b } \kappa \right ) \cdot \left \{ \begin{array}{l} \exp \left ( - i \frac{a}{b} \kappa \right ) \\ \exp \left ( i \frac{a}{b} \kappa \right ) \end{array} \right. .
\end{equation}

We can substitute this into the RHS of the equation for $( l_0^2 )_{em}$ (\ref{DDl0^2em=(Dphi)^2}) and find $( l_0^2 )_{em}$ by acting by the discrete propagator $(\sum_j \oDelta_j \Delta_j)^{-1}$ on this RHS. Namely, if
\begin{equation}                                                           
\sum_j \oDelta_j \Delta_j ( l_0^2 )_{em} = f ( \bom )
\end{equation}

\noindent is a function on the vertices $\bom$, then, passing to the momentum representation,
\begin{eqnarray}\label{l0^2em=(DD)^{-1}f}                                  
& & \hspace{-10mm} ( l_0^2 )_{em} ( \bx ) = \frac{1}{2} \iint\limits^{{} ~~~ \pi \,\, \pi}_{- \pi -\pi} \frac{\d^2 \bl}{(2 \pi )^2} \exp \left ( i l_1 \frac{x_1 }{b } + i l_2 \frac{x_2 }{b } \right ) \int\limits^\pi_{- \pi } \frac{\d l_3 }{2 \pi } \frac{ \exp ( i l_3 k_3 )}{ \ch \kappa ( l_1, l_2 ) - \cos l_3 } \nonumber \\ & & \cdot \sum_{ \bom } e^{- i \bl \bom } f ( \bom ) , ~~~ k_3 = x_3 / b .
\end{eqnarray}

In $f ( \bom )$, double and quadruple products of $\phi$, $\ophi$ appear. Taking them in the form of the integrals (\ref{phi,ophi}) over $\d q_{j 1} \d q_{j 2}$, $j = 1, 2$ or $j = 1, 2, 3, 4$, we get integral combinations of the exponentials $\prod_j \exp ( i q_{j 1} m_1 + i q_{j 2} m_2 - | m_3 | \kappa_j )$, $\kappa_j = \kappa ( q_{j 1}, q_{j 2} )$. Then the summation over $\bom$ giving the Fourier transform of $f ( \bom ) $ in (\ref{l0^2em=(DD)^{-1}f}) leads to
\begin{eqnarray}\label{sum_m3exp(im3-|m3|)}                             
\sum_{m_a} \exp [ i m_a ( \sum_j q_{j a} - l_a )] = 2 \pi \delta ( \sum_j q_{j a} - l_a ) , ~~~ a = 1, 2 , \\ \sum_{m_3} \exp ( - i l_3 m_3 - | m_3 | \sum_j \kappa_j ) = \frac{ \sh \sum_j \kappa_j}{ \ch \sum_j \kappa_j - \cos l_3} .
\end{eqnarray}

\noindent The integration over $\d l_3$ of the resulting dependence on $l_3$ in (\ref{l0^2em=(DD)^{-1}f}), (\ref{sum_m3exp(im3-|m3|)}) can be performed in a closed form,
\begin{eqnarray}                                                           
\int \frac{\exp ( i l_3 k_3 ) \sh \sum_j \kappa_j }{ [ \ch \kappa ( l_1, l_2 ) - \cos l_3 ] ( \ch \sum_j \kappa_j - \cos l_3 ) } \frac{ \d l_3 }{ 2 \pi } = \frac{1 }{ \ch \sum_j \kappa_j - \ch \kappa ( l_1, l_2 ) } \nonumber \\ \cdot \left ( \frac{ \sh \sum_j \kappa_j }{\sh \kappa ( l_1, l_2 )} e^{ - \kappa ( l_1 , l_2 ) | k_3 |} - e^{ - \sum_j \kappa_j | k_3 | } \right ) .
\end{eqnarray}

\noindent The delta-functions (\ref{sum_m3exp(im3-|m3|)}) allow to eliminate the integration over $\d q_{j 1} \d q_{j 2}$ for one of the two-dimensional quasi-momenta $\bq_j$ expressing it through the other $\bq_j$'s and $\bl$.

As a result, $( l_0^2 )_{em}$ is a sum of expressions with factors $\exp ( - i a \sum_j \epsilon_j \kappa_j / b )$ under the integral sign, $j = 1, 2$ or $j = 1, 2, 3, 4$, $\epsilon_j = \pm 1$. Let us denote these expressions as $I^{ \{ \epsilon_j \} }$ where $\{ \epsilon_j \}$ is a brief notation for the sequence of signs of $\epsilon_j$. We have
\begin{eqnarray}\label{1loop-general}                                      
& & ( l_0^2 )_{em} = I^{++} + I^{+-} + I^{++++} , \mbox{ where for } \{ \epsilon_j \} = ++ \mbox{ or } +- \nonumber \\ & & I^{\{\epsilon_j\}} = \frac{G Q^2}{b^2 } \Re \iint \frac{\d^2 \bl}{(2 \pi )^2} \exp \left ( i l_1 \frac{x_1 }{b } + i l_2 \frac{x_2 }{b } \right ) \iint \frac{\d^2 \bq_1 }{\sh \kappa_1 \sh \kappa_2 } \nonumber \\ & & \cdot \exp \left ( - i \frac{a}{b} \sum_j \epsilon_j \kappa_j \right ) \frac{ \sh \kappa_1 \sh \kappa_2 - \sin q_{11} \sin q_{21} - \sin q_{12} \sin q_{22} }{ \ch \sum_j \kappa_j - \ch \kappa ( l_1, l_2 ) } \nonumber \\ & & \cdot \left ( \frac{ \sh \sum_j \kappa_j }{\sh \kappa ( l_1, l_2 )} e^{ - \kappa ( l_1 , l_2 ) | k_3 |} - e^{ - \sum_j \kappa_j | k_3 | } \right ) .
\end{eqnarray}

\noindent In the case $j = 1, 2, 3, 4$, it is useful to write out an expression for the general set $\{ \epsilon_j \}$, which contains the case $\{ \epsilon_j \} = ++++$ of interest here,
\begin{eqnarray}\label{3loop-general}                                      
\hspace{-5mm} I^{\{\epsilon_j\}} & = & - \frac{G Q^2}{b^2 } \Re \iint \frac{\d^2 \bl}{(2 \pi )^2} \exp \left ( i l_1 \frac{x_1 }{b } + i l_2 \frac{x_2 }{b } \right ) \iint \iint \iint \frac{\prod^\prime_j \d^2 \bq_j}{(2 \pi )^2 \prod_j \sh \kappa_j} \nonumber \\ & & \cdot \frac{\exp ( - i a \sum_j \epsilon_j \kappa_j / b )}{ \ch \sum_j \kappa_j - \ch \kappa ( l_1, l_2 ) } \left ( \frac{ \sh \sum_j \kappa_j }{\sh \kappa ( l_1, l_2 )} e^{ - \kappa ( l_1 , l_2 ) | k_3 |} - e^{ - \sum_j \kappa_j | k_3 | } \right )
\end{eqnarray}

\begin{figure}[b]
\centerline{\includegraphics[width=9.0cm]{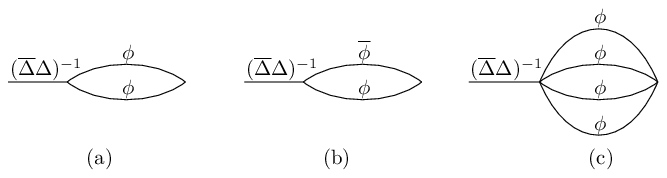}}
\caption{The diagrams describing the procedure of finding $( l_0^2 )_{em}$ from (\ref{DDl0^2em=(Dphi)^2}). \label{f1}}
\end{figure}

\noindent ($\prod^\prime_j$ means $\prod_j$ with some one of $j$ omitted). Schematically, this can be considered as the result of some transformation of a contribution of 1-loop and 3-loop diagrams to which the expression of $( l_0^2 )_{em}$ from (\ref{DDl0^2em=(Dphi)^2}) corresponds (Fig.~\ref{f1}). The number of these diagrams (three) is equal to the number of terms having different degrees $n_+$, $n_-$ of dependence on $\phi$, $\ophi$ of the type $\phi^{n_+} (\ophi )^{n_-}$ in $T_{0 0} $ or on the RHS of equation (\ref{DDl0^2em=(Dphi)^2}) for $( l_0^2 )_{em}$. We can identically (on the Kerr-Newman solution) rewrite the continuum expression for $T_{0 0} $ and get another set of such terms when passing to the finite-difference form. It seems quite probable that the number of such terms cannot be less than three. The present set, consisting of three terms, is singled out by a well-defined expansion over coordinates and powers of $b$, as considered in Conclusion.

\subsection{Saddle Point Estimate}\label{saddle}

It is convenient to pass from $q_{j 1}, q_{j 2}$ to variables $\alpha_j , \beta_j$ according to
\begin{equation}\label{q_j=al+-bl}                                         
\left. \begin{array}{rcl} q_{j 1} & = & \alpha_j l_1 - \beta_j l_2  \\ q_{j 2} & = & \alpha_j l_2 + \beta_j l_1 \end{array} \right \} ,
\end{equation}

\noindent then, due to $\delta^2 (\sum_j \bq_j - \bl )$ above,
\begin{equation}                                                           
\sum_j \alpha_j = 1, ~~~ \sum_j \beta_j = 0 .
\end{equation}

\begin{figure}[b]
\centerline{\includegraphics[width=9.0cm]{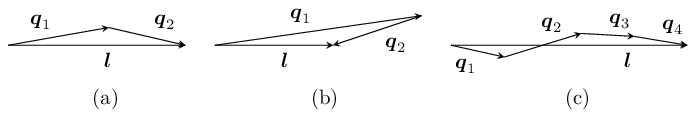}}
\caption{Configurations in the neighborhood of the saddle point $\bq_j \propto \bl \forall j$ (or $\beta_j = 0 \forall j$ in (\ref{q_j=al+-bl})) for $\kappa_1 + \kappa_2$ (a), $\kappa_1 - \kappa_2$ (b), $\kappa_1 + \kappa_2 +\kappa_3 + \kappa_4$ (c) for the diagrams (a), (b), (c) of Fig.~\ref{f1}, respectively. \label{f2}}
\end{figure}

There are exponential factors under the integral sign of the type $\exp ( i \dots / b )$, which can be considered when $b$ is small. Let us use the saddle point method. Of such factors, the dependence on $\alpha_j , \beta_j$ is contained in $\exp ( - i a \sum_j \epsilon_j \kappa_j / b )$. In the first approximation (temporarily omitting lattice-inspired non-linearities) $\kappa_j$ is the length of $\bq_j$, $\kappa_j = ( q_{j 1}^2 + q_{j 2}^2 )^{1 / 2}$. Using the triangle inequalities, it is easy to see that a saddle point over all $\beta_j$'s is achieved at $\beta_j = 0$ $\forall$  $j$ (Fig.~\ref{f2}).

In the actual calculation, the lattice corrections to $( q_{j 1}^2 + q_{j 2}^2 )^{1 / 2}$ can be taken into account to get genuine $\kappa_j$ and terms of the expansion around $\{ \beta_j | \beta_j = 0 \forall j \}$ can be considered. Namely, we denote for quasi-momenta
\begin{equation}\label{sin_l=L_cos_varphi_sin_varphi}                      
\sin \frac{l_a}{2} = \lambda_a , ~ a = 1, 2, \mbox{  and  } \left. \begin{array}{rcl} \lambda_1 & = & \Lambda \cos \varphi  \\ \lambda_2 & = & \Lambda \sin \varphi \end{array} \right \}
\end{equation}

\noindent and for the observation point coordinates
\begin{equation}                                                           
\left. \begin{array}{rcr} x_1 & = & r_0 \cos \varphi_0 + k_1 b  \\ x_2 & = & r_0 \sin \varphi_0 + k_2 b \\ x_3 & = & k_3 b \end{array} \right \} .
\end{equation}

\noindent That is, we consider an $O ( b )$-neighborhood of a point $\bx = ( r_0 \cos \varphi_0 , r_0 \sin \varphi_0 , 0 ) $ chosen to ensure an enhancement in $l_0^2 ( \bx )$ analogous to the infinite enhancement on the singularity ring in the continuum.

Let us write out a few first terms of the expansions of the expressions of interest over $\beta_j , \Lambda$.
\begin{eqnarray}\label{kappa_j}                                            
& & \kappa_j = 2 \Lambda \sqrt{\alpha_j^2 + \beta_j^2 } + \frac{1}{3} | \alpha_j | \Lambda^3 \left ( 1 - \frac{1}{2} \sin^2 2 \varphi \right ) - \frac{1}{3} | \alpha_j |^3 \Lambda^3 \left ( 2 - \frac{1}{2} \sin^2 2 \varphi \right ) + \dots , \nonumber \\ & & \sqrt{\alpha_j^2 + \beta_j^2 } = | \alpha_j | + \frac{\beta_j^2}{2 | \alpha_j |} + \dots , \nonumber \\ & & l_1 x_1 + l_2 x_2 = 2 \Lambda r_0 \cos ( \varphi - \varphi_0 ) + \frac{1}{3} r_0 \Lambda^3 ( \cos \varphi_0 \cos^3 \varphi + \sin \varphi_0 \sin^3 \varphi ) + \dots , \nonumber \\ & & \kappa ( l_1 , l_2 ) = 2 \Lambda - \frac{1}{3} \Lambda^3 + \dots .
\end{eqnarray}

\noindent Looking ahead, $\beta_j$, $\Lambda$ are typically of order $O( b^{1 / 3} )$, and $\alpha_j = O( 1 )$, and the terms up to $O( b )$ are written out, and those of order $O( b^{4 / 3} )$ are disregarded; although, further these orders of magnitude of the variables will be overridden in special cases so that the whole series of such terms will need to be summed. We have $\exp [ i b^{- 1} ( l_1 x_1 + l_2 x_2 - a \sum_j \epsilon_j \kappa_j ) ]$ under the integral sign where
\begin{equation}                                                           
l_1 x_1 + l_2 x_2 - a \sum_j \epsilon_j \kappa_j = 2 \Lambda \left [ r_0 \cos ( \varphi - \varphi_0 ) - a \sum_j \epsilon_j | \alpha_j | \right ] + \dots
\end{equation}

\noindent (in the lowest order in $\beta_j$, $\Lambda$).

Thus, $\varphi = \varphi_0$ is one else characterization of the saddle point, and this is achieved if we choose $r_0=a$ and the region $\sgn \alpha_j = \epsilon_j$ so that $r_0 \cos ( \varphi - \varphi_0 ) - a \sum_j \epsilon_j | \alpha_j | = O ((\varphi - \varphi_0 )^2 )$ due to $\sum_j \alpha_j = 1$. Thus, the former singularity ring $r_0=a$ is the locus of points at the nearest vertices to which $( l_0^2 )_{em}$ (as well as $( l_0^2 )_g$ and $A_0$) should grow substantially.

Then we have in the vicinity of the saddle point
\begin{eqnarray}                                                           
& & l_1 x_1 + l_2 x_2 - a \sum_j \epsilon_j \kappa_j = - a \Lambda (\varphi - \varphi_0 )^2 - a \Lambda \sum_j \frac{ \beta_j^2 }{ \alpha_j } + \frac{1}{3} \mu a \Lambda^3 \sum_j \alpha_j^3 + \dots , \nonumber \\ & & \mu = 1 + \sin^4 \varphi_0 + \cos^4 \varphi_0 = \frac{7}{4} + \frac{1}{4} \cos 4 \varphi_0 .
\end{eqnarray}

The integration element in the variables $\Lambda$, $\varphi$, $\alpha_j$, $\beta_j$ takes the form
\begin{eqnarray}                                                           
& & \d^2 \bl \d^2 \bq_1 = 2^4 \Lambda^3 \d \Lambda \d \varphi \d^2 \beta \delta \left (\sum\nolimits_j \beta_j \right ) \d^2 \alpha \delta \left (\sum\nolimits_j \alpha_j - 1 \right ) , \nonumber \\ & & \d^2 \bl \d^2 \bq_1 \d^2 \bq_2 \d^2 \bq_3 = 2^8 \Lambda^7 \d \Lambda \d \varphi \d^4 \beta \delta \left (\sum\nolimits_j \beta_j \right ) \d^4 \alpha \delta \left (\sum\nolimits_j \alpha_j - 1 \right ) .
\end{eqnarray}

The saddle point integration over $\beta_j$, $\varphi$ can be performed,
\begin{eqnarray}\label{dvarphiDbeta}                                       
& & \int \exp \left ( - i \frac{a}{b} \Lambda \sum\nolimits_j \frac{\beta_j^2 }{\alpha_j } - i \frac{a}{b} \Lambda \varphi^2 \right ) \delta \left (\sum\nolimits_j \beta_j \right ) \d \varphi \prod\nolimits_j \d \beta_j \nonumber \\ & & = \int \exp \left ( - i \frac{a}{b} \Lambda \sum\nolimits_j \frac{\beta_j^2 }{\alpha_j } - i \frac{a}{b} \Lambda \varphi^2 + i \nu \sum\nolimits_j \beta_j \right ) \d \varphi \frac{\d \nu }{2 \pi } \prod\nolimits_j \d \beta_j \nonumber \\ & & = \left \langle \beta_j \Rightarrow \beta_j + \frac{b \nu }{2 a \Lambda } \alpha_j \right \rangle = \nonumber \\ & & = \int \exp \left ( - i \frac{a}{b} \Lambda \sum\nolimits_j \frac{\beta_j^2 }{\alpha_j } - i \frac{a}{b} \Lambda \varphi^2 + i \frac{b }{4 a \Lambda } \nu^2 \right ) \d \varphi \frac{\d \nu }{2 \pi } \prod\nolimits_j \d \beta_j \nonumber \\ & & = \prod\nolimits_j \left [ \exp \left ( - i \sgn \alpha_j \frac{\pi }{4 } \right ) \sqrt{\frac{\pi b}{a \Lambda} | \alpha_j | } \right ]
\end{eqnarray}

\noindent ($\varphi - \varphi_0$ is redesignated as $\varphi$).

It is easy to get that the introduction of the additional factor $\beta_j^2$ or $\varphi^2$ under the integral sign will lead to the additional factor
\begin{equation}\label{<beta2><varphi2>}                                   
\beta_j^2 \Rightarrow \langle \beta_j^2 \rangle = \frac{i }{2 } \frac{b }{ a \Lambda } ( \alpha_j^2 - \alpha_j ) \mbox{ ~ or ~ } (\varphi - \varphi_0)^2 \Rightarrow \langle (\varphi - \varphi_0)^2 \rangle = - \frac{i }{2 } \frac{b }{ a \Lambda }
\end{equation}

\noindent on the RHS of (\ref{dvarphiDbeta}) (a kind of vacuum average). At small $\Lambda$, they show an inaccuracy of the saddle point approximation when integrating over $\beta_j$, $\varphi$. For $\langle \beta_j^2 \rangle$ for small $\Lambda$, the effective $\beta_j^2$ violates the requirement $| \beta_j | < | \alpha_j |$, which is necessary for the convergence of the expansion of $( \alpha_j^2 + \beta_j^2 )^{1 / 2}$ in (\ref{kappa_j}) over $\beta_j^2$. For $\langle (\varphi - \varphi_0)^2 \rangle$, an infinite growth follows for $\Lambda \to 0$, while $\varphi $ changes in a compact region. These circumstances can lead to a divergence at $\Lambda \to 0$ when calculating the contributions due to the terms of the expansion of the integrand over $\beta_j$, $\varphi$ in sufficiently high orders. Therefore, consider cutting off the integration by the condition
\begin{equation}\label{L>b}                                                
\Lambda \geq \frac{b }{\zeta a } ,
\end{equation}

\noindent where $\zeta $ is a parameter of the order of 1.

Note that in what follows, when the regularization is removed ($\zeta \to \infty$), only a certain contribution to $l_0^2$ (the structure 1) diverges, and this divergence is logarithmic. Imposing the condition (\ref{L>b}), we can consider that for $\Lambda < b / (\zeta a )$ the (compact) integration over $\varphi$ gives a constant ($2 \pi$), just as for $\Lambda = 0$. Therefore, the subsequent integration over $\Lambda$ in the region $\Lambda < b / (\zeta a )$ already converges, and its contribution comes with an additional small factor $b$ and can be discarded in the main order over $b$.

Returning to the expressions for $I^{ \{ \epsilon_j \} }$, we denote $k_{1 2} = k_1 \cos \varphi_0 + k_2 \sin \varphi_0$. The leading order over $\beta_j , \Lambda$ is taken and the expansion in powers of $| k_3 |$ up to $k_3^2$ is made. We have for $j = 1, 2$ (below $\d \alpha$ is $\d \alpha_1$ or $\d \alpha_2$)
\begin{eqnarray}                                                           
& & I^{ \{ \epsilon_j \} } = \frac{4 G Q^2}{\pi b a} \Re \iint \Lambda^2 \d \Lambda \d \alpha \sqrt{ | \alpha_1 \alpha_2 | } \frac{ | \alpha_1 \alpha_2 | - \alpha_1 \alpha_2 }{ | \alpha_1 \alpha_2 | } \nonumber \\ & & \cdot \left [ \frac{ 1 }{ 2 \Lambda^2 ( \sum_j | \alpha_j | + 1 ) } - \frac{2 }{3 } \Lambda \sum\nolimits_j | \alpha_j | | k_3 | - \frac{ \sum_j | \alpha_j | }{ \sum_j | \alpha_j | + 1 } k_3^2 \right ] \nonumber \\ & & \cdot \exp \left ( i \frac{ a }{ b } \frac{ \mu }{ 3 } \Lambda^3 \sum\nolimits_j \alpha_j^3 + 2 i k_{1 2} \Lambda - i \frac{\pi }{ 4 } \sum\nolimits_j \epsilon_j \right ) ,
\end{eqnarray}

\noindent and for $j = 1, 2, 3, 4$ ($\d^3 \alpha \equiv \prod^\prime_j \d \alpha_j$)
\begin{eqnarray}\label{I-3loop-1order}                                     
& & I^{ \{ \epsilon_j \} } = - \frac{ G Q^2 }{ \pi^2 a^2 } \Re \iiiint \frac{ \Lambda \d \Lambda \d^3 \alpha }{ \prod_j \sqrt{ | \alpha_j | }} \nonumber \\ & & \cdot \left [ \frac{ 1 }{ 2 \Lambda^2 ( \sum_j | \alpha_j | + 1 ) } - \frac{2 }{3 } \Lambda \sum\nolimits_j | \alpha_j | | k_3 | - \frac{ \sum_j | \alpha_j | }{ \sum_j | \alpha_j | + 1 } k_3^2 \right ] \nonumber \\ & & \cdot \exp \left ( i \frac{ a }{ b } \frac{ \mu }{ 3 } \Lambda^3 \sum\nolimits_j \alpha_j^3 + 2 i k_{1 2} \Lambda - i \frac{\pi }{ 4 } \sum\nolimits_j \epsilon_j \right ) .
\end{eqnarray}

To single out the dependence on $b$, we rescale $\Lambda$ to get a new variable $w$ instead of it,
\begin{equation}\label{L=w/alpha}                                          
\Lambda = w \left ( \frac{3 b}{ \mu a \sum_j \alpha_j^3 } \right )^{1 / 3} .
\end{equation}

\noindent Then we have for $j = 1, 2$
\begin{eqnarray}\label{I2loop-w-xi}                                        
& & I^{ \{ \epsilon_j \} } = \frac{12 G Q^2}{\pi \mu a^2 } \Re \iint w^2 \d w \d \alpha \frac{ \sqrt{ | \alpha_1 \alpha_2 | } }{ \sum_j \alpha_j^3 } \frac{ | \alpha_1 \alpha_2 | - \alpha_1 \alpha_2 }{ | \alpha_1 \alpha_2 | } \nonumber \\ & & \cdot \left [ \frac{1 }{2 } \left ( \frac{ \mu a }{ 3 b } \right )^{ 2 / 3 } \frac{1 }{ w^2 } \frac{ ( \sum_j \alpha_j^3 )^{ 2 / 3 } }{ \sum_j | \alpha_j | + 1 } - \frac{2 }{3 } \left ( \frac{3 b}{ \mu a } \right )^{1 / 3} w \frac{ \sum_j | \alpha_j | }{ ( \sum_j \alpha_j^3 )^{ 1 / 3 } } | k_3 | - \frac{ \sum_j | \alpha_j | }{ \sum_j | \alpha_j | + 1 } k_3^2 \right ] \nonumber \\ & & \cdot \exp \left [ i w^3 + 2 i k_{1 2} \left ( \frac{3 b}{ \mu a } \right )^{1 / 3} \frac{ w }{ ( \sum_j \alpha_j^3 )^{ 1 / 3 } } - i \frac{\pi }{ 4 } \sum\nolimits_j \epsilon_j \right ] ,
\end{eqnarray}

\noindent and for $j = 1, 2, 3, 4$
\begin{eqnarray}                                                           
& & I^{ \{ \epsilon_j \} } = - \frac{ G Q^2 }{ \pi^2 a^2 } \left ( \frac{3 b}{ \mu a } \right )^{2 / 3} \Re \iiiint \frac{ w \d w \d^3 \alpha }{ ( \sum_j \alpha_j^3 )^{ 2 / 3 } \prod_j \sqrt{ | \alpha_j | }} \nonumber \\ & & \cdot \left [ \frac{1 }{2 } \left ( \frac{ \mu a }{ 3 b } \right )^{ 2 / 3 } \frac{1 }{ w^2 } \frac{ ( \sum_j \alpha_j^3 )^{ 2 / 3 } }{ \sum_j | \alpha_j | + 1 } - \frac{2 }{3 } \left ( \frac{3 b}{ \mu a } \right )^{1 / 3} w \frac{ \sum_j | \alpha_j | }{ ( \sum_j \alpha_j^3 )^{ 1 / 3 } } | k_3 | - \frac{ \sum_j | \alpha_j | }{ \sum_j | \alpha_j | + 1 } k_3^2 \right ] \nonumber \\ & & \cdot \exp \left [ i w^3 + 2 i k_{1 2} \left ( \frac{3 b}{ \mu a } \right )^{1 / 3} \frac{ w }{ ( \sum_j \alpha_j^3 )^{ 1 / 3 } } - i \frac{\pi }{ 4 } \sum\nolimits_j \epsilon_j \right ] .
\end{eqnarray}

\subsection{Expansion Over Coordinates}\label{expansion}

We see that $I^{++}$ in the first approximation is zero, since $| \alpha_1 \alpha_2 | - \alpha_1 \alpha_2 = 0$ in the corresponding region $\alpha_1 > 0$, $\alpha_2 > 0$. The value $( | \alpha_1 \alpha_2 | - \alpha_1 \alpha_2 ) / | \alpha_1 \alpha_2 |$ is $( \sh \kappa_1 \sh \kappa_2 - \sin q_{11} \sin q_{21} - \sin q_{12} \sin q_{22} ) / ( \sh \kappa_1 \sh \kappa_2 )$ in the first approximation, and to get a non-zero result, it is necessary to take into account the subsequent terms of the expansion over $\beta_j , \Lambda$. These terms lead to an additional smallness for small $b$ in comparison with $I^{+-}$. Thus, $I^{++}$ can be discarded unless $I^{+-}$ accidentally turns out to be equal to zero in the first approximation. In addition, $I^{++++}$ has an additional smallness for small $b$ compared to $I^{+-}$, again, unless $I^{+-}$ happens to be equal to zero in the first approximation.

Thus, $I^{+-}$ usually dominates. Unlike $I^{++}$ and $I^{++++}$, it contains integration over $\alpha_j$ in an infinite domain, and there may be divergences. They depend on the considered structure from those structures into which $I^{+-}$ is decomposed,
\begin{equation}                                                           
I^{+-} = \sum\nolimits_{p_1, p_2, p_3} I^{+-}_{k_1^{p_1} k_2^{p_2} | k_3 |^{p_3}} , ~~~ I^{+-}_{k_1^{p_1} k_2^{p_2} | k_3 |^{p_3}} = {\rm const} \cdot k_1^{p_1} k_2^{p_2} | k_3 |^{p_3} .
\end{equation}

\noindent Namely, it is seen that $I^{+-}_1$, $I^{+-}_{| k_3 |}$ and $I^{+-}_{ k_3^2 }$ contain divergent integrations over $\d \alpha$. The contributions of the remaining structures, namely $k_1^{p_1} k_2^{p_2} | k_3 |^{p_3}$, where either $p_1 \geq 1$ and/or $p_2 \geq 1$ and/or $p_3 \geq 3$, contain additional factors $\Lambda$ under the integral sign and, thus, additional negative powers of $\sum_j \alpha_j^3$ (see (\ref{L=w/alpha})), which improve the convergence of the integral over $\d \alpha$. As a result, this integral over $\d \alpha$ converges for these contributions. For example, consider such a structure $k_{1 2}$ of the minimal total degree $p_1 + p_2 + p_3$ (which is equal to 1). We denote
\begin{equation}                                                           
\alpha_1 = 1 + \xi , ~~~ \alpha_2 = - \xi
\end{equation}

\noindent and find
\begin{eqnarray}                                                           
& & \hspace{-0mm} I^{+-}_{k_{1 2}} = \frac{12 G Q^2}{\pi a^2 \mu } \left ( \frac{ \mu a }{ 3 b } \right )^{1 / 3} k_{1 2} \Re \int\nolimits_0^\infty i e^{\textstyle i w^3 } w \d w \int\nolimits_0^\infty \frac{\sqrt{\xi^2 + \xi } \d \xi }{(3 \xi^2 + 3 \xi + 1)^{2 / 3} ( \xi + 1 )} \nonumber \\ & & \hspace{-0mm} = - \left ( \frac{ 4 }{ 3 } \right )^{1 / 3} \frac{ \Gamma ( 1 / 3 ) }{ \Gamma ( 2 / 3 ) } \left [ 1 - \frac{ 3 }{ 2^{1 / 3} \pi } \frac{ \Gamma^3 ( 2 / 3 ) }{ \Gamma^2 ( 1 / 3 ) } \, {}_2 \! F_1 \left ( \frac{ 1 }{ 2 } , \frac{ 1 }{ 2 } ; \frac{ 7 }{ 6 } ; - \frac{ 1 }{ 3 } \right ) \right ] \frac{ G Q^2 }{ \mu^{2 / 3} a^{5 / 3} b^{ 1 / 3}} k_{1 2} \nonumber \\ & & = - ( 1.64 ... ) \cdot \frac{ G Q^2 }{ \mu^{2 / 3} a^{5 / 3} b^{ 1 / 3}} k_{1 2} .
\end{eqnarray}

Products of similar integrals over $\d w$ and $\d \xi $ follow for the structures $k_{1 2} | k_3 |$ and $k_{1 2}^2$. As can be read from (\ref{I2loop-w-xi}), the normal order of magnitude for $I^{+-}_{k_{1 2} | k_3 |}$ is $O ( b^{ 2 / 3 } )$. For $I^{+-}_{k_{1 2}^2}$, such an order is $O ( 1 )$, but the integral over $\d w$ is zero for it, $\Re \int_0^\infty \exp ( i w^3 ) w^2 \d w = 0$. Thus, the actual order of magnitude for $k_{1 2}^2$ is determined by $I^{++}_{k_{1 2}^2}$, $I^{++++}_{k_{1 2}^2}$ and subsequent orders of expansion of the integrand in $I^{+-}_{k_{1 2}^2}$ over $\Lambda$, $\beta_j$ and $\varphi - \varphi_0$ and turns out to be equal to $O ( b^{ 2 / 3 } )$  as well.
\begin{equation}                                                           
I^{+-}_{k_{1 2} | k_3 |} = O ( b^{ 2 / 3 } ) k_{1 2} | k_3 |, ~~~ I^{ \{ \epsilon_j \} }_{k_{1 2}^2} = O ( b^{ 2 / 3 } ) k_{1 2}^2 .
\end{equation}

\noindent This means a smallness compared to the contribution to these structures in $( l_0^2 )_g$, where it is $O ( b^{ 1 / 3 } )$.

As for the divergent integrations over $\alpha_j$ (in $I^{+-}_1$, $I^{+-}_{| k_3 |}$, $I^{+-}_{ k_3^2 }$), if $\Lambda$ is given, $\alpha_j$ are actually bounded from above, since $q_{j 1}$, $q_{j 2}$ are bounded as quasi-momentum components,
\begin{equation}\label{aj+-bj<pi}                                          
\left. \begin{array}{rcl} | q_{j 1} | & \leq & \pi \\ | q_{j 2} | & \leq & \pi \end{array} \right \} \Rightarrow \left. \begin{array}{rcl} | 2 \alpha_j \arcsin ( \Lambda \cos \varphi_0 ) - 2 \beta_j \arcsin ( \Lambda \sin \varphi_0 ) | & \leq & \pi \\ | 2 \alpha_j \arcsin ( \Lambda \sin \varphi_0 ) + 2 \beta_j \arcsin ( \Lambda \cos \varphi_0 ) | & \leq & \pi \end{array} \right \}
\end{equation}

\noindent (using definitions (\ref{q_j=al+-bl}), (\ref{sin_l=L_cos_varphi_sin_varphi})). Since we are expanding over $\beta_j$, here $\beta_j$ should be considered small, and we neglect it\footnote{If we were asking about such $\alpha_j$ that $\exists$ $\beta_j$ such that equations (\ref{aj+-bj<pi}) hold, then we would get $| \alpha_j | \leq \pi (| \cos \varphi_0 | + | \sin \varphi_0 | )/(2 \Lambda )$, which is less restrictive than (\ref{aj<pi}) (although it coincides with it in some important points, $\varphi_0 = 0$ and $\varphi_0 = \pi / 4$).}. Taking $\Lambda$ small, we find
\begin{equation}\label{aj<pi}                                              
| \alpha_j | \leq \frac{ \tau }{ \Lambda } , ~~~ \tau = \frac{ \pi }{ 2 {\rm max} \{ | \cos \varphi_0 | , | \sin \varphi_0 | \}} .
\end{equation}

In the expansion of the integrand over $\Lambda$, the coefficient in front of each power of $\Lambda$ grows for large $\xi$ like $\xi$ to a power not exceeding the power of $\Lambda$. The above integration limit $\xi \leq \tau / \Lambda$ (\ref{aj<pi}) admits large $\xi$ for small $\Lambda$, and the contribution of such $\Lambda$, $\xi$ is mainly due to the terms $\propto (\Lambda \xi )^p$ in the expansion. We should neglect the terms proportional to $\Lambda^p$ times $\xi$ to a power less than $p$. In other words, we come back to (\ref{1loop-general}) and, after the saddle point integration, consider the integrand in the limit $| \bl | \ll | \bq_j |$ or $\xi \gg 1$.

In this way, we have for the considered structures
\begin{equation}\label{I1+I|k3|+Ik32}                                      
I^{+-}_{1(0)} + I^{+-}_{| k_3 | (0)} + I^{+-}_{ k_3^2 (0) } = \frac{4 G Q^2}{ \pi a b } \Re \iint \Lambda^2 \d \Lambda \xi \d \xi \left ( \frac{1}{ \Lambda } f_1 + f_{| k_3 |} | k_3 | + f_{ k_3^2 } k_3^2 \right ) e^{ i W } ,
\end{equation}

\noindent where the subscript $(0)$ means that some further corrections to these values are to be analyzed; $ f_1 $, $ f_{| k_3 |}$, $ f_{ k_3^2 }$ are some functions of $\Lambda \xi \equiv q$ obtained by substituting $\bq_j$, for example,
\begin{equation}                                                           
\left. \begin{array}{l} q_{1 1} \\ q_{1 2} \end{array} \right \} = 2 ( \xi + 1 ) \arcsin \left \{ \begin{array}{l} \Lambda \cos \varphi_0 \\ \Lambda \sin \varphi_0 \end{array} \right. = 2 q \left \{ \begin{array}{l} \cos \varphi_0 \\ \sin \varphi_0 \end{array} \right.
\end{equation}

\noindent at $\xi \gg 1$, $\Lambda \ll 1$ into the integrand. They are given in \ref{appendix1}. And for the exponent $W$, we have the difference between the values of $\kappa$ for two quasi-momenta $\bq_1$ and $\bq_2$ close in absolute value,
\begin{eqnarray}                                                           
W = \frac{ a }{ b } \Lambda \varpi ( q ) , ~~~ \varpi & = & 2 - \lim_{\Lambda \to 0} \frac{ \kappa ( q_{1 1}, q_{1 2} ) - \kappa ( q_{2 1}, q_{2 2} ) }{ \Lambda } \nonumber \\ & = & 2 - \frac{ \partial }{ \partial q } \kappa ( 2 q \cos \varphi_0 , 2 q \sin \varphi_0 ) \stackrel{q \to 0}{ \to } \mu q^2 .
\end{eqnarray}

We can pass from $\Lambda$, $\xi$ to $q$, $W$ as new variables,
\begin{equation}                                                           
\left. \begin{array}{rcl} q & = & \Lambda \xi \\ W & = & \frac{a }{ b } \Lambda \varpi (\Lambda \xi ) \end{array} \right \} , ~~~ \left. \begin{array}{rcl} \Lambda & = & \frac{b }{ a } \frac{W }{ \varpi ( q ) } \\ \xi & = & \frac{ a }{ b } \frac{ q \varpi (q ) }{ W } \end{array} \right \} , ~~~ \d \Lambda \d \xi = \d q \frac{ \d W }{ W } .
\end{equation}

\noindent In these variables, the integration region
\begin{equation}                                                           
1 \leq \xi \leq \frac{ \tau }{ \Lambda } , \mbox{ where } \frac{ b }{ a \zeta } \leq \Lambda \leq \tau ,
\end{equation}

\noindent takes the form
\begin{equation}                                                           
\frac{ 1 }{ \zeta } \varpi ( q ) \leq W \leq \frac{ a }{ b } q \varpi ( q ) , \mbox{ where } \frac{ b }{ a \zeta } \leq q \leq \tau
\end{equation}

\begin{figure}[b]
\centerline{\includegraphics[width=7.0cm]{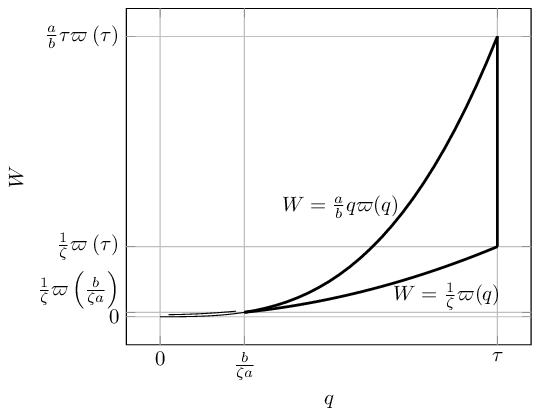}}
\caption{The figure shows the region of integration (circled by a thick line). \label{f3}}
\end{figure}

\noindent (Fig.~\ref{f3}).

Then we obtain for the structure 1
\begin{eqnarray}\label{I1(0)}                                              
I^{+-}_{1 (0)} = \frac{4 G Q^2}{ \pi a b } \Re \left [ \int_{ \frac{1}{ \zeta } \varpi \left ( \frac{ b }{ \zeta a } \right )}^{ \frac{1}{ \zeta } \varpi ( \tau )} e^{i W} \frac{\d W}{ W } \int_{\tilde{\varpi}^{-1} \left ( \frac{ b }{ a } W \right )}^{ \varpi^{ - 1 } ( \zeta W ) } q f_1 ( q ) \d q \right. \nonumber \\ \left. + \int_{ \frac{1}{ \zeta } \varpi \left ( \tau \right )}^{ \frac{a}{ b } \tau \varpi ( \tau )} e^{i W} \frac{\d W}{ W } \int_{\tilde{\varpi}^{-1} \left ( \frac{ b }{ a } W \right )}^\tau q f_1 ( q ) \d q \right ] \nonumber \\ \stackrel{\stackrel{\scriptstyle \zeta \to \infty}{b \to 0}}{ = } \frac{4 G Q^2}{ \pi a b } \left [ \int_0^\tau \frac{ \d \varpi ( x ) }{ \varpi ( x ) } \int_0^x q f_1 ( q ) \d q + \int_{ \frac{ \varpi ( \tau ) }{ \zeta }}^\infty \frac{ \cos W }{ W } \d W \int_0^\tau q f_1 ( q ) \d q \right ] \nonumber \\ = \frac{4 G Q^2}{ \pi a b } \left \{ \int_0^\tau \left [ \ln \frac{\varpi (\tau)}{ \varpi ( q ) } \right ] q f_1 ( q ) \d q + \left ( \ln \frac{ \zeta }{ \varpi ( \tau ) } - \gamma \right ) \int_0^\tau q f_1 ( q ) \d q \right \} \nonumber \\ = \frac{2 G Q^2}{ \pi a b } \tau \left ( 2 + \ln \frac{ \zeta }{ \varpi ( \tau ) } - \gamma \right ) ,
\end{eqnarray}

\noindent where $\tilde{\varpi} (q) \equiv q \varpi(q)$, and $\varpi^{-1}$, $\tilde{\varpi}^{-1}$ mean inverse functions. In the last equality, we use $\varpi = \mu q^2$ and $f_1 = ( 2 q )^{- 1}$ for a rough estimate.

For control, we consider corrections to the saddle point method, when the terms linear in $\beta_j^2$ and $\varphi^2$ are taken into account not only in the exponent, but also in the rest of the integrand. In \ref{beta+varphi}, (\ref{I-beta2-phi2}), these terms are written out. Integrating over $\beta_j$ and $\varphi$ by substituting $\langle \beta_j^2 \rangle$ and $\langle (\varphi - \varphi_0)^2 \rangle$ (\ref{<beta2><varphi2>}) there, taking large $| \alpha_j |$ (or $\xi$) and passing to the variables $q$, $W$, we find the transformed and corrected (\ref{I1+I|k3|+Ik32}) ($(0) + (\beta^2) + (\varphi^2)$ terms) in the minimum order over $\Lambda$ or $q$,
\begin{eqnarray}\label{0+b2+varphi2}                                       
I^{+-}_1 + I^{+-}_{| k_3 |} + I^{+-}_{ k_3^2 } = \frac{4 G Q^2}{ \pi a b } \Re \iint \d q \d W \left \{ \frac{1}{2W} \left [ 1 - \frac{ i }{4} \frac{ \varpi ( q ) }{W} - \frac{i}{ 3 } \frac{q^2 \varpi (q)}{ W } \cos 4 \varphi_0 \right ] \right. \nonumber \\ - \frac{3}{8} \frac{ b }{a} \frac{ q^2 }{ \varpi (q) } \left[ 1 + \frac{ i }{4} \frac{ \varpi ( q ) }{W} - 2 \frac{i}{ 3 } \frac{q^2 \varpi (q)}{ W } \cos 4 \varphi_0 \right ] | k_3 | \nonumber \\ \left. - 2 \frac{ b }{a} \frac{ q }{ \varpi (q) } \left [ 1 + \frac{ i }{8} \frac{ b }{ a q } - \frac{i}{ 2 } \frac{q^2 \varpi (q)}{ W } \cos 4 \varphi_0 \right ] k_3^2 \right \} e^{ i W } . ~~~
\end{eqnarray}

\noindent Here, the $\beta_j^2$ and $\varphi^2$ corrections are singled out by the different $i$-ness of the integrand compared to the main term. The $\varphi^2$ correction is also distinguished by the factor $\cos 4 \varphi_0$.

Modifying the integrand in (\ref{I1(0)}) according to (\ref{0+b2+varphi2}), we obtain $(\beta^2) + (\varphi^2)$ corrections for the structure 1,
\begin{equation}                                                           
I^{+-}_{1, (\beta^2) + (\varphi^2)} = \frac{ G Q^2 \mu }{ \pi a b } \left [ \frac{\tau^3 }{ 6 } \left ( \frac{ 5 }{ 3 } + \ln \frac{ \zeta }{ \varpi ( \tau ) } - \gamma \right ) + \frac{ 2 \tau^5 }{ 15 } \left ( \frac{ 7 }{ 5 } + \ln \frac{ \zeta }{ \varpi ( \tau ) } - \gamma \right ) \cos 4 \varphi_0 \right ] .
\end{equation}

\noindent It has a numerical smallness compared to $I^{+-}_{1 (0)}$. Besides, averaging $I^{+-}_{1 (\varphi^2)}$ (but not its square) over the lattice orientation yields zero due to $\cos 4 \varphi_0$. $\zeta$ is assumed to be large, but, on the other hand, there is no large parameter here, and there is nowhere to come from with numerical greatness. Therefore, it is natural to choose $\zeta$ so that $\ln [ \zeta / \varpi (\tau ) ] \simeq 1$; then the order of magnitude estimate reads
\begin{equation}                                                           
I^{+-}_1 \simeq \frac{4 G Q^2}{ \pi a b } \tau .
\end{equation}

Then for the structures $| k_3 |$, $k_3^2$ we obtain
\begin{eqnarray}                                                           
& & \left. \begin{array}{l} I^{+-}_{ | k_3 | (0) }  \\ I^{+-}_{ k_3^2 (0) } \end{array} \right \}  = \frac{4 G Q^2}{ \pi a^2 } \Re \int_{\frac{ b }{ a \zeta }}^\tau \frac{ q \d q }{ \varpi (q) } \left \{ \begin{array}{l} f_{ | k_3 | } ( q )  \\ f_{ k_3^2 } ( q ) \end{array} \right \} \int_{\frac{\varpi (q) }{\zeta }}^{\frac{a}{b} q \varpi (q) } e^{i W} \d W \left \{ \begin{array}{l} | k_3 |  \\ k_3^2 \end{array} \right. \nonumber \\ & & = \frac{4 G Q^2}{ \pi a^2 } \int_{\frac{ b }{ a \zeta }}^\tau \frac{ q \d q }{ \varpi (q) } \left \{ \begin{array}{l} f_{ | k_3 | } ( q )  \\ f_{ k_3^2 } ( q ) \end{array} \right \} \left \{ \sin \left [ \frac{a}{b} q \varpi (q) \right ] - \sin \frac{\varpi (q)}{\zeta } \right \} \left \{ \begin{array}{l} | k_3 |  \\ k_3^2 \end{array} \right. \nonumber \\ & & \stackrel{\stackrel{\scriptstyle \zeta \to \infty}{b \to 0}}{ = } \frac{4 G Q^2}{ 3 \pi a^2 \mu } \int_0^\infty \frac{\sin x}{x} \d x \left \{ \begin{array}{l} f_{ | k_3 | } ( 0 ) | k_3 |  \\ f_{ k_3^2 } ( 0 ) k_3^2 \end{array} \right.  = - \frac{4}{3} \frac{G Q^2}{a^2 \mu} \left \{ \begin{array}{l} 0 \cdot | k_3 |  \\ k_3^2 \end{array} \right. .
\end{eqnarray}

\noindent This result is determined only by $f_{| k_3 |}$, $f_{ k_3^2 }$ at $q = 0$ and by the behaviour of $\varpi (q)$ at $q \to 0$. Strictly speaking, we get $I^{+-}_{ | k_3 | (0) } = O ( b^{1 / 3} )$ for $f_{| k_3 |} \propto q$ at small $q$, regardless of the detailed form of $f_{| k_3 |}$, which is zero in the considered order $O ( 1 )$. At the same time, the $(\beta^2) + (\varphi^2)$ corrections turn out to contribute to $O ( 1 )$.

The $(\beta^2) + (\varphi^2)$ corrections for these structures, $I^{+-}_{|k_3|, (\beta^2) + (\varphi^2)}$ and $I^{+-}_{k_3^2, (\beta^2) + (\varphi^2)}$, are easier to find by integrating first over $\d q$, then over $\d W$, again as in (\ref{I1(0)}), changing the integrand according to (\ref{0+b2+varphi2}). The correction $I^{+-}_{k_3^2 (\beta^2)}$ can be considered zero, as a higher order in $b$, since it contains $b$ as a factor in (\ref{0+b2+varphi2}). With the $(0)$-terms found, the result reads
\begin{eqnarray}                                                           
I^{+-}_{|k_3|} & = & \frac{4}{9} \frac{G Q^2}{ a^2 } \tau^3 \left ( 1 - \frac{8}{5} \tau^2 \cos 4 \varphi_0 \right ) | k_3 | , \nonumber \\ I^{+-}_{k_3^2} & = & - \frac{4}{3} \frac{G Q^2}{ a^2 } \left ( 1 + \frac{3}{8} \tau^4 \cos 4 \varphi_0 \right ) k_3^2 .
\end{eqnarray}

Thus, the result reads
\begin{eqnarray}\label{l0em2}                                              
& & \hspace{-5mm} ( l_0^2 )_{em} = \frac{4 G Q^2}{ \pi a b } \tau - ( 1.64 ... ) \cdot \frac{ G Q^2 }{ \mu^{2 / 3} a^{5 / 3} b^{ 1 / 3}} k_{1 2} + \frac{4}{9} \frac{G Q^2}{ a^2 } \tau^3 \left ( 1 - \frac{8}{5} \tau^2 \cos 4 \varphi_0 \right ) | k_3 | \nonumber \\ & & + O ( b^{ 2 / 3 } ) k_{1 2}^2 + O ( b^{ 2 / 3 } ) k_{1 2} | k_3 | - \frac{4}{3} \frac{G Q^2}{ a^2 } \left ( 1 + \frac{3}{8} \tau^4 \cos 4 \varphi_0 \right ) k_3^2 .
\end{eqnarray}

\noindent This should be added to the earlier found \cite{Kha2}
\begin{eqnarray}                                                           
( l_0^2 )_g & = & \frac{r_g}{a\sqrt{\mu}} \frac{\Gamma(1 / 3)}{\Gamma(2 / 3)} \left ( \frac{a \mu}{6 b} \right )^{1/3} - \frac{2 r_g}{a \sqrt{3\mu}} |k_3| + \nonumber \\ & + & \frac{r_g}{a\sqrt{\mu}} \frac{\Gamma(2 / 3)}{\Gamma(1 / 3)} \left ( \frac{6 b}{a \mu} \right )^{1/3} \left ( -k_{12}^2 + k_3^2 +  \frac{2}{\sqrt{3}} k_{12} | k_3 | \right ) .
\end{eqnarray}

The ratio of the em-term to the g-term includes a large factor $O( b^{- 2 / 3})$ at the point $\bk = {\bf 0}$ (i.e., on the ring $r_0 = a$) and has the same (positive) sign as the ratio at the center in the Reissner-Nordström solution, where we have found \cite{our3} that it includes a large factor $O( b^{- 1})$ and has a positive sign,
\begin{eqnarray}                                                           
( l_0^2 )_{em} ({\bf 0}) : ( l_0^2 )_g ({\bf 0}) \propto \frac{G Q^2}{a b} : \frac{r_g}{a^{2 / 3} b^{1 / 3}} \mbox{(Kerr-Newman)} \nonumber \\ \mbox{vs } \frac{G Q^2}{b^2} : \frac{r_g}{b} \mbox{(Reissner-Nordström)} .
\end{eqnarray}

\noindent This differs from that at macroscopic distances or in the continuum case (where it tends to infinity), where such a ratio is negative.

The ratio of the bilinear in $\bk$ part of $(l_0^2)_{em}$ to the bilinear part of $(l_0^2)_g$, taken as the ratio of the coefficients at $k_3^2$, includes a large factor $ O ( b^{- 1 / 3} )$,
\begin{equation}\label{k32-em-to-k32-g}                                    
\frac{ \mbox{coefficient at } k_3^2 \mbox{ in } ( l_0^2 )_{em} }{ \mbox{coefficient at } k_3^2 \mbox{ in } ( l_0^2 )_g } = - \frac{2^{5 / 3} \Gamma(1 / 3)}{3^{4 / 3} \Gamma(2 / 3)} \frac{G Q^2 \mu^{5 / 6}}{r_g a^{2 / 3} b^{1 / 3}} = - ( 2.3 ... ) \cdot \frac{G Q^2}{r_g a^{2 / 3} b^{1 / 3}} .
\end{equation}

\noindent Here we have taken $\mu = 7 / 4 + (\cos 4 \varphi_0 ) / 4$ as its average value $7 / 4$. This ratio characterizes (outside the sources, which condition certainly holds for $k_3\neq 0$) the ratio of the em- and g-parts of the components of the discrete Riemann tensor $R_{\lambda \mu \nu \rho}$. For the main contribution at $b \to 0$ to the latter we can write in the order $[l ]^2$ the same formula that we have found \cite{Kha2} for the pure Kerr geometry,
\begin{eqnarray}\label{Rlmnr+R3l3m-on-ring}                                
& & R_{\lambda \mu \nu \rho} = \frac{1}{ 2 b^2 } \left [ - n_\lambda n_\rho \oDelta_{ ( \mu } \Delta_{ \nu ) } ( l_0^2 ) - n_\mu n_\nu \oDelta_{ ( \lambda } \Delta_{ \rho ) } ( l_0^2 ) \right. \nonumber \\ & & \left. + n_\mu n_\rho \oDelta_{ ( \lambda } \Delta_{ \nu ) } ( l_0^2 ) + n_\lambda n_\nu \oDelta_{ ( \mu } \Delta_{ \rho ) } ( l_0^2 ) \right ] , ~ \oDelta_{ ( \lambda } \Delta_{ \mu ) } \equiv \frac{1}{2} \left ( \oDelta_\lambda \Delta_\mu + \oDelta_\mu \Delta_\lambda \right ) ,
\end{eqnarray}

\noindent where we can substitute $l_0^2 = ( l_0^2 )_g + ( l_0^2 )_{em}$. In reality, there are arguments \cite{Wald} that it would be very difficult for any astrophysical body to achieve and/or maintain a charge to mass ratio of greater than $\sim 10^{- 18}$ (in "geometrized" units, $G = c = 1$). If we substitute instead of the charge its upper bound $10^{- 18}$ of the mass according to these arguments, the modulus of the ratio (\ref{k32-em-to-k32-g}) becomes
\begin{equation}                                                           
5.8 \cdot 10^{- 37} \frac{r_g }{ a^{2 / 3} b^{1 / 3}} .
\end{equation}

\noindent As considered in Section \ref{method} (the paragraph following that one with (\ref{int-S D-Omega=F})), when constructing the formalism, it is important that the ratio $l_{\rm Pl} / b$ be a small parameter (that is, the dimensionless parameter $\eta$ would be large, $\eta \gg 1$) or $b \gg l_{\rm Pl}$. So we take $b = 10 l_{\rm Pl} = 8 \cdot 10^{- 32} cm$ as the lower bound. Even for black holes with the largest known $r_g$ ($\sim 10^{19 } cm$) that probably exist in the Universe, this ratio reaches 1 only at microscopic sizes $a$ ($\sim 10^{- 10 } cm$), i.e. for a practically non-rotating black hole, which seems to be an unlikely event. That is, the electromagnetic contribution to the curvature on the former singularity ring of a black hole rotating in any noticeable way is usually relatively small.

\section{Conclusion}

Here, when analyzing the metric in the discrete version of the Kerr-Newman solution, we get a kind of discrete diagram technique with internal electromagnetic lines with a finite number (several) of such diagrams. A simplifying circumstance is the three-dimensionality (static character) of the diagrams, a complicating circumstance is the use of a complex extension of a discrete static electromagnetic propagator (similar to the Newman-Janis complex extension in the continuum case).

We can note the following three features of the present calculation and its result.

First, we note some uniqueness of the discrete equation for the metric function $l_0^2$ (\ref{DDl0^2em=(Dphi)^2}) obtained by the identical transformation of the RHS $\propto T_{00}$ (\ref{4piTlm}) on the Kerr-Newman type solution.

Such an original equation for $l_0^2$ with a bilinear RHS $\propto T_{00}$ in $\phi , \ophi$, where $\phi$ is the Coulomb/Newton potential extended to complex coordinates, also has an explicit dependence on coordinates, which makes the transition to a discrete form ambiguous, especially where this dependence is strong (near the former singularity ring). This is a dependence on two coordinates (because of the axial symmetry), and we can go to two new coordinates $\Re \phi$, $\Im \phi$. Thus, the RHS can be identically (on the Kerr-Newman solution) transformed to a form without an explicit coordinate dependence, but with monomials in $\phi$, $\ophi$ of degree greater than two.

As we then find, such a general monomial $\phi^{n_+} (\ophi )^{n_-}$ leads to a diagram for $l_0^2$, for which the expansion over the integer coordinates $k_1, k_2, k_3$ is at the same time an expansion over positive powers of $b^{1 / 3}$ with coefficients that are integrals over $\d^{n_+ + n_- -1} \alpha_j = \delta (\sum_j \alpha_j - 1) \d^{n_+ + n_-} \alpha_j $ of expressions with negative powers of $( \sum_j \alpha_j^3 )^{1 / 3}$, where $n_+$ variables $\alpha_j$ are positive, while $n_-$ variables $\alpha_j$ are negative. Meanwhile, $\sum_j \alpha_j^3$ is strictly separated from zero only if either all $\alpha_j$ are positive, or there is a negative $\alpha_j$, but $j = 1, 2$ ($n_+ + n_- = 2$); otherwise, $\sum_j \alpha_j^3$ can pass through zero, and there is a complication in integrating over $\d^{n_+ + n_- -1} \alpha_j $ in high-order terms of this series expansion. For example, if we restrict ourselves to rewriting only the third term in $T_{0 0} $ as $\propto (\Re \phi )^2(\Im \phi )^2$, then we get $\phi^2(\ophi )^2$ on the RHS and the complication when expanding over $k_1, k_2, k_3$, as just mentioned.

Thus, the accepted transition from $T_{0 0}$ (\ref{4piTlm}) through the sum of terms $\propto ( \bpart \Re \phi )^2$ and $\propto \Re \left ( \phi^4 \right )$ (in the continuum) to the discrete form in (\ref{DDl0^2em=(Dphi)^2}) is singled out by a well-defined expansion over coordinates and powers of $b$. This is done here to further show that the contribution in the leading order over metric variations comes from the discrete form of $( \bpart \Re \phi )^2$ or $(\bpart \phi )(\bpart \ophi )$ on the RHS (except when the contribution to some structures accidentally vanishes in the leading order).

Second, the expansion over coordinates in the vicinity of the former singularity ring consists of a regular part, which receives its contribution mainly from quasi-momenta of order $O( b^{1 / 3} )$, and an irregular part (three first terms), to which a small quasi-momentum $O ( b )$ and a loop quasi-momentum close to the maximum $O ( 1 )$ make the main contribution.

Third, the result (\ref{l0em2}) for the electromagnetic contribution to the metric function $( l_0^2 )_{em}$ indeed has structures with coefficients formally infinitely larger at $b \to 0$ than the coefficients at these structures in $( l_0^2 )_g$. For the structure 1, the contributions to $( l_0^2 )_{em}$ and $( l_0^2 )_g$ have the same sign, contrary to what one might expect, assuming an analogy with the behaviour of these quantities when approaching the singularity ring in the continuum theory. Earlier, we have found the same equality of signs of $( l_0^2 )_{em}$ and $( l_0^2 )_g$ at the center in the discrete Reissner-Nordström solution. The ratio of the contribution of $( l_0^2 )_{em}$ to the Riemann tensor to the contribution of $( l_0^2 )_g$ formally tends to infinity at $b \to 0$ as $b^{ - 1 / 3}$. However, assuming a natural upper bound on the possible charge of ordinary physical bodies and taking into account that $b$ cannot be arbitrarily small in the present approach ($b \gg l_{\rm Pl}$), we find that the electromagnetic part of the Riemann tensor on the former singularity ring is relatively small for possible black holes found in the Universe. Meanwhile, a non-rotating (Reissner-Nordström) black hole may well have a dominant electromagnetic part of the metric and of the Riemann tensor at the center.

\appendix

\section{The integrand at large loop quasi-momenta}\label{appendix1}

The functions defining the integrand in (\ref{I1+I|k3|+Ik32}) are
\begin{eqnarray}                                                           
& & f_1 = \frac{1}{2} \frac{\sh^2 \kappa_1 + \sin^2 q_{1 1} + \sin^2 q_{1 2}}{\sh^2 \kappa_1} \frac{\ch \kappa_1}{\sh \kappa_1} \nonumber \\ & & = \frac{1}{2} \frac{(s^2_c + s^2_s + s^2_c s^2_s) ( 1 + 2 s^2_c + 2 s^2_s )}{(s^2_c + s^2_s)^{3 / 2} ( 1 + s^2_c + s^2_s )^{3 / 2}} \stackrel{q \to 0}{ \to } \frac{ 1 }{ 2 q } , \nonumber \\ & & f_{| k_3 |} = \frac{\sh^2 \kappa_1 + \sin^2 q_{1 1} + \sin^2 q_{1 2}}{\sh^2 \kappa_1} \frac{ \kappa_1 - \sh \kappa_1 \ch \kappa_1 }{ \sh^2 \kappa_1 } \nonumber \\ & & = \frac{ \arsh [ 2 (s^2_c + s^2_s )^{1 / 2} (1 + s^2_c + s^2_s )^{1 / 2} ] - 2 (s^2_c + s^2_s )^{1 / 2} (1 + s^2_c + s^2_s )^{1 / 2} (1 + 2 s^2_c + 2 s^2_s ) }{2 (s^2_c + s^2_s )^2 (1 + s^2_c + s^2_s )^2 } \nonumber \\ & & \cdot (s^2_c + s^2_s + s^2_c s^2_s) \stackrel{q \to 0}{ \to } - \frac{ 8 }{ 3 } q , \nonumber \\ & & f_{ k_3^2 } = - \frac{\sh^2 \kappa_1 + \sin^2 q_{1 1} + \sin^2 q_{1 2}}{\sh^2 \kappa_1} \frac{ \kappa_1^2}{\sh^2 \kappa_1} \nonumber \\ & &  = - \frac{\arsh^2 [ 2 (s^2_c + s^2_s )^{1 / 2} (1 + s^2_c + s^2_s )^{1 / 2} ]}{2 (s^2_c + s^2_s )^2 (1 + s^2_c + s^2_s )^2} (s^2_c + s^2_s + s^2_c s^2_s) \stackrel{q \to 0}{ \to } -2 .
\end{eqnarray}

\noindent Here
\begin{equation}                                                           
\left. \begin{array}{l} s_c \\ s_s \end{array} \right \} \equiv \sin \left \{ \begin{array}{l} q \cos \varphi_0 \\ q \sin \varphi_0 \end{array} \right. .
\end{equation}

\section{$\beta^2$-, $\varphi^2$-corrections to the saddle point approximation}\label{beta+varphi}

Here we expand up to the terms $\beta_j^2$ and $\varphi^2$ not only in the exponent, but also in the rest of the integrand. Besides that, the expansion over $\Lambda$ is implied.

The products of terms linear in $\varphi$ in two different factors under the integral also lead to terms proportional to $(\varphi - \varphi_0)^2$, but of a higher order in $\Lambda$, and such terms in the given consideration are not taken into account (and they have a numerical smallness in the coefficients).
\begin{eqnarray}\label{I-beta2-phi2}                                       
& & I^{ \{ \epsilon_j \} } = \frac{4 G Q^2}{\pi b a} \Re \iint \Lambda^2 \d \Lambda \d \alpha \sqrt{ | \alpha_1 \alpha_2 | } \left [ \frac{ | \alpha_1 \alpha_2 | - \alpha_1 \alpha_2 }{ | \alpha_1 \alpha_2 | } + \frac{ \langle \beta^2 \rangle }{ 2 | \alpha_1 \alpha_2 | \alpha_1 \alpha_2 } \right. \nonumber \\
& & \left. - \langle (\varphi - \varphi_0)^2 \rangle \Lambda^2 \frac{ \alpha_1 \alpha_2 }{| \alpha_1 \alpha_2 |} (\alpha_1^2 + \alpha_2^2) \cos 4 \varphi_0 \right ] \left \{ \frac{ 1 }{ 2 \Lambda^2 ( \sum\limits_j | \alpha_j | + 1 ) } \right. \nonumber \\
& & \cdot \left [ 1 - \frac{1}{2 (\sum\limits_j | \alpha_j | + 1)} \sum_j \frac{\langle \beta_j^2 \rangle}{ | \alpha_j | } + \frac{\Lambda^2 \langle (\varphi - \varphi_0)^2 \rangle \cos 4 \varphi_0 }{3 (\sum\limits_j | \alpha_j | + 1 )} \sum_j | \alpha_j | ( 1 - \alpha_j^2) \right ] \nonumber \\
& & - \frac{2 }{3 } \Lambda \sum_j | \alpha_j | | k_3 | \nonumber \\
& & \cdot \left [ 1 + \frac{1}{2 \sum\limits_j | \alpha_j | } \sum_j \frac{\langle \beta_j^2 \rangle}{ | \alpha_j | } - \frac{\Lambda^2 \langle (\varphi - \varphi_0)^2 \rangle \cos 4 \varphi_0 }{3 \sum\limits_j | \alpha_j | } \sum_j | \alpha_j | ( 1 - \alpha_j^2) \right ] \nonumber \\
& & - \frac{ \sum_j | \alpha_j | }{ \sum\limits_j | \alpha_j | + 1 } k_3^2 \cdot \nonumber \\
& & \hspace{-7mm} \left. \left [ 1 + \frac{1}{2 ( \sum\limits_j | \alpha_j | + 1 ) \sum\limits_j | \alpha_j | } \sum_j \frac{\langle \beta_j^2 \rangle}{ | \alpha_j | } - \frac{\Lambda^2 \langle (\varphi - \varphi_0)^2 \rangle \sum\limits_j | \alpha_j | ( 1 - \alpha_j^2) }{3 ( \sum\limits_j | \alpha_j | + 1 ) \sum\limits_j | \alpha_j | } \cos 4 \varphi_0 \right ] \right \} \nonumber \\
& & \cdot \exp \left ( i \frac{ a }{ b } \frac{ \mu }{ 3 } \Lambda^3 \sum\nolimits_j \alpha_j^3 + 2 i k_{1 2} \Lambda - i \frac{\pi }{ 4 } \sum\nolimits_j \epsilon_j \right ) ,
\end{eqnarray}

\noindent As written, this expression contains a bilinear part in $\beta_j^2$, $(\varphi - \varphi_0)^2$, but only the linear part in them should be considered. A similar expression for $I^{ \{ \epsilon_j \} }$ for the 3-loop diagram is obtained by literally replacing the contents in square brackets in (\ref{I-3loop-1order}) with the contents in curly brackets in (\ref{I-beta2-phi2}).

\section*{Acknowledgments}

The present work was supported by the Ministry of Education and Science of the Russian Federation.

\end{document}